\documentclass[twocolumn]{aastex631}

\usepackage{tabularx}

\usepackage{upquote}
\usepackage{threeparttable}
\usepackage{amsmath} 
\usepackage{soul}

\begin{document}

\title{Systematic Analysis of Changing-look AGN Variability Using ZTF Light Curves}

\correspondingauthor{Huimei Wang}

\author[0000-0001-8803-0738]{Huimei Wang}
\email{wang\_hm@pku.edu.cn}
\affil{Department of Astronomy, School of Physics, Peking University, Beijing 100871, People's Republic of China}
\affil{Kavli Institute for Astronomy and Astrophysics, Peking University, Beijing 100871, People's Republic of China}

\author[0000-0002-7350-6913]{Xue-Bing Wu}
\email{wuxb@pku.edu.cn}
\affil{Department of Astronomy, School of Physics, Peking University, Beijing 100871, People's Republic of China}
\affil{Kavli Institute for Astronomy and Astrophysics, Peking University, Beijing 100871, People's Republic of China}

\author{Nanyu Yao}
\affil{Department of Astronomy, School of Physics, Peking University, Beijing 100871, People's Republic of China}

\author[0000-0001-8879-368X]{Bing Lyu}
\affil{Department of Astronomy, School of Physics, Peking University, Beijing 100871, People's Republic of China}
\affil{Kavli Institute for Astronomy and Astrophysics, Peking University, Beijing 100871, People's Republic of China}

\author[0009-0005-3823-9302]{Yuxuan Pang}
\affil{Department of Astronomy, School of Physics, Peking University, Beijing 100871, People's Republic of China}
\affil{Kavli Institute for Astronomy and Astrophysics, Peking University, Beijing 100871, People's Republic of China}

\author[0000-0002-0759-0504]{Yuming Fu}
\affil{Leiden Observatory, Leiden University, Einsteinweg 55, 2333 CC Leiden, The Netherlands}
\affil{Kapteyn Astronomical Institute, University of Groningen, P.O. Box 800, 9700 AV Groningen, The Netherlands}

\author[0000-0002-0792-2353]{Rui Zhu}
\affil{Department of Astronomy, School of Physics, Peking University, Beijing 100871, People's Republic of China}
\affil{Kavli Institute for Astronomy and Astrophysics, Peking University, Beijing 100871, People's Republic of China}

\author[0000-0002-6893-3742]{Qian Yang}
\affil{Center for Astrophysics | Harvard \& Smithsonian, 60 Garden Street, Cambridge, MA 02138, USA}

\begin{abstract}

Changing-look active galactic nuclei (CLAGNs) are a unique population of AGNs that exhibit the appearance (turn-on) or disappearance (turn-off) of broad emission lines. This study aims to explore the intrinsic mechanisms of CLAGNs by investigating their photometric variability using data from the Zwicky Transient Facility (ZTF), which has provided high-cadence observations over the past five years. By visual inspections, we construct a sample of 152 CLAGNs from the literature, all of which show spectral transitions and large optical variability in their ZTF light curves. By analyzing 90 of these CLAGNs and the control samples of Type 1 AGNs, Type 2 AGNs, and extremely variable quasars (EVQs), matched in redshift ($0.2<z<0.8$) and supermassive black hole mass, we compare the color variability, structure function (SF), and variability metric $\sigma_{\mathrm{QSO}}$, which quantifies how closely the light curves resemble a damped random walk (DRW) model. We find that while CLAGNs and EVQs differ from typical Type 1 and Type 2 AGNs in bolometric luminosity and Eddington ratio, the on/off-state CLAGNs share similar variability patterns with the overall CLAGN population, and distinct from EVQ, Type 1 and Type 2 AGNs. 
This suggests that 'on' and 'off' CLAGNs are not simply equivalent to Type 1 and Type 2 AGNs, respectively. Instead of undergoing genuine transitions between two AGN types, CLAGNs may inhabit a critical state where moderate fluctuations in accretion rate lead to the temporary spectral changes.


\end{abstract}

\section{Introduction} \label{sec:intro}
Active Galactic Nuclei (AGNs) are among the most luminous and dynamic objects in the Universe, powered by accretion onto super-massive black holes (SMBHs) at the centers of galaxies. AGNs are classified into Type 1 and Type 2 AGNs based on their emission line features. Type 1 AGNs show broad and narrow emission lines, while Type 2 AGNs only show narrow emission lines. In the unified model of AGNs, the difference in viewing angle affect the obscuration of the broad line region (BLR) by the dusty torus and cause the difference between Type 1 and Type 2 AGNs \citep{Urry1995}.

Changing-look active galactic nuclei (CLAGNs) refer to a special type of AGNs that exhibit significant spectral changes, characterized by the appearance or disappearance of broad emission lines. Most of these objects also exhibit large amplitude changes in luminosity \citep{MacLeod2019}. Over the past decades, with the accumulation of large-scale spectroscopic survey data, the number of identified CLAGNs has increased significantly, with approximately 1000 CLAGNs discovered to date \citep{Yang2018,MacLeod2019, Sheng2019, PottsVillforth2021, Jin2022, Yang2023, Guo2024, Guo2025, Dong2025}. It is generally believed that the spectral type transitions in CLAGNs are primarily driven by three mechanisms: (1) tidal disruption events \citep{vanVelzen2020}, (2) obscuration by intervening clouds \citep{Mereghetti2021}, and (3) intrinsic changes in accretion rate \citep{LaMassa2015,MacLeod2016,Graham2020,Yang2023}. The majority of CLAGN transitions are thought to be caused by intrinsic variations in the accretion rate \citep{Ricci2023}, although the exact physical mechanism behind these changes remains unclear.

The variability is a signature of quasars across all wavelengths and timescales. In the rest-frame UV and optical bands, the continuum emission from the accretion disc of quasars typically varies about 0.2 mag at a time scale of days to years \citep{VB2004,MacLeod2012}. \citet{MacLeod2019} found that CLAGNs have larger variability than typical AGNs. \citet{Ren2022} found that extremely variable quasars (EVQs), with large optical photometric variability, lie in the tail of a broad distribution of quasar properties, and are not a distinct population, similar to CLAGNs, though EVQs do not exhibit the dramatic emergence or disappearance of broad emission lines.
However, it is yet unclear whether the variability in CLAGNs and normal AGNs are driven by the same mechanisms. 

While many studies have focused on the spectral transitions of CLAGNs, understanding their variability in photometric data, such as color and magnitude changes, is equally important \citep{MacLeod2012}. Photometric variability studies can reveal additional insights about the accretion process and the nature of transition between Type 1 to Type 2 states. \citet{Kovacevic2025} compares the optical variability of Seyfert 1 and Seyfert 2 AGNs in the Swift BAT sample using TESS and ASAS-SN light curves. It was found that Seyfert 1s exhibit significantly stronger variability, supporting the orientation-based unification models.
Thus, it is worth building large samples of CLAGNs and EVQs, to explore whether they have different variability properties from normal Type 1 and Type 2 AGNs.

In this study, we plan to investigate the color variability, structure function, and variability metric of CLAGNs using data from the Zwicky Transient Facility (ZTF). ZTF is an ongoing time-domain photometric survey started in 2018, designed to monitor the entire northern sky. This survey utilizes a wide-field camera mounted on the Samuel Oschin Telescope at Palomar Observatory, and provide public data releases (DR) that offer relatively deep photometric light curves, with a g-band limiting magnitude of 20.8 mag \citep{Masci2019}. ZTF is notable for its high cadence, capturing observations approximately every 3 days, which is crucial for studying the rapid variability of astronomical objects, including a large sample of AGNs \citep{WangShu2024}.

We compile a sample of CLAGNs and a corresponding control sample composed of Type 1 AGNs, Type 2 AGNs, and EVQs, matched in redshift and black hole mass. By comparing CLAGNs with Type 1 and Type 2 AGNs, as well as EVQs, we aim to determine whether the variability mechanisms in CLAGNs are distinct from those of other AGNs or they share common origins.

This paper is organized as follows. We describe the data and sample selection in Section \ref{sec:data}. Results and discussions are given in Section \ref{sec:result} and \ref{sec:discussion}, respectively. Throughout this paper, we adopt a standard $\Lambda \mathrm{CDM}$ cosmology with $\Omega_{\Lambda}$ = 0.7, $\Omega_m$ = 0.3, and $H_0 = 70\, \mathrm{km\, s^{-1}\, Mpc^{-1}}$.

\section{Data} \label{sec:data}
\subsection{ZTF lightcurve}\label{sec:ZTF}
We extract ZTF DR23 \footnote{https://www.ztf.caltech.edu/ztf-public-releases.html} light curves for quasars in our final sample in the g-band and r-band. The ZTF DR23 has light curves up to 6 years in length with yearly gaps. We also exclude bad measurements where the \texttt{catflags} parameter is not zero, and remove significant outliers that deviate by more than $3\sigma$ from the local median. We also perform a 3-day binning to reduce scatter. 

\begin{figure}[htbp]
    \centering
    \includegraphics[width=0.45\textwidth]{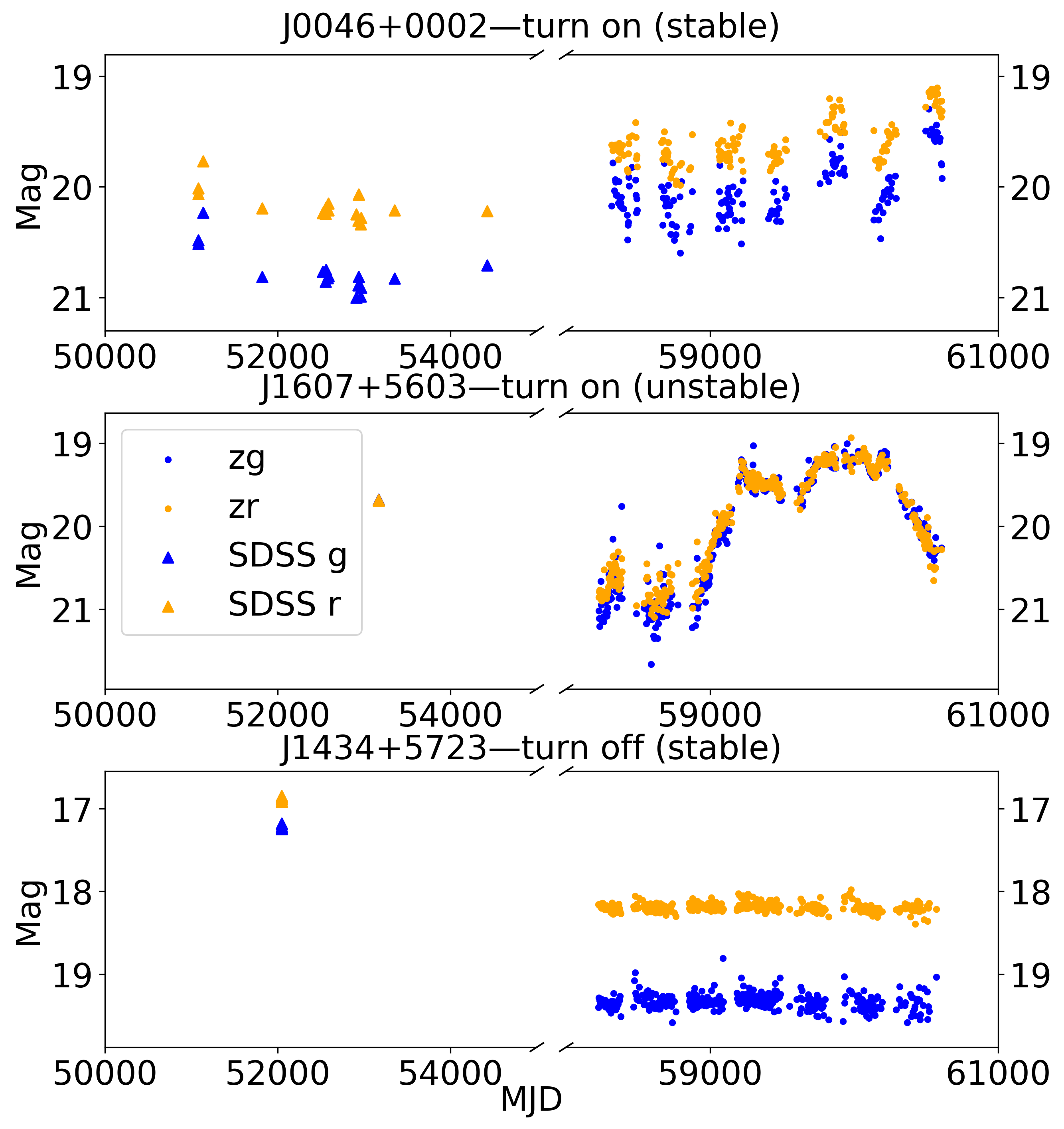}
    \caption{Light curves of three CLAGN sources in different states: (a) turn-on and recently stable, (b) turn-on and recently unstable, (c) turn-off and recently stable. ZTF $g$- and $r$-band measurements are shown as blue and orange points, while SDSS measurements are shown as blue and orange triangles, respectively. The legend in (b) is the same for (a) and (c).}
    \label{fig:clagn_lightcurves}
\end{figure}

\subsection{Changing-look AGN Sample}\label{sec:CLAGNs}

To date, the number of optical CLAGNs found based on various methods is about 300. There are mainly two methods to select CLAGNs: 1) repeat spectra (e.g. Sloan Digital Sky Survey (SDSS, \citet{Paris2018}), Dark Energy Spectroscopic Instrument spectra (DESI, \citet{DESI2025}), Large Sky Area Multi-Object Fiber Spectroscopic Telescope (LAMOST, \citet{Cui2012})) in different or multi-epoch spectral surveys \citep{Lopez2023,PottsVillforth2021,Yang2023,Zeltyn2024}; 2) photometric variability and follow-up spectroscopic observations \citep{Yang2018, Sheng2019, MacLeod2019, Yang2024}. Both of these method often involve visual inspection as the last step. 
Table \ref{Table:photometry} lists some typical photometric and spectroscopic criteria to find CLAGN candidates.

\begin{table*}[htb!]
\centering
\small
\begin{threeparttable}
\caption{Overview of CLAGN searching. The upper part show the sources found by photometric change, and the lower part show the sources found by spectroscopic change. For the same series of papers finding CLAGNs, we only show the latest work and the criterion.}\label{Table:photometry}

\begin{tabularx}{\textwidth}{ccc}
\hline
\textbf{Reference} & \textbf{Band}   & \textbf{Photometric Selection Criteria}  \\
\hline
\cite{Yang2018}&       Optical + Infrared        &   turn on: $\Delta W1 \textless -0.2$ and $\Delta (W1 - W2)> 0.1$; $\Delta g\textless 0$ and $g\textless19$ \textsuperscript{a}
       \\
& &turn off: $\Delta W1> 0.2$ and $\Delta (W1 - W2) \textless -0.1$; $\Delta g> 1$ \\
\cite{MacLeod2019}&Optical&$|\Delta g| > 1$ mag, $|\Delta r| > 0.5$ mag ($\sigma \textless$  0.15 mag)\textsuperscript{b}\\
\cite{Sheng2019}& Infrared&$\Delta W1 > 0.4$ and $\Delta W2 > 0.4$ at $> 5 \sigma$ significance, \\
&&$\mathrm{max}(W1-W2)> 0.4$ and $\mathrm{min}(W1-W2)\textless 0.8$\\

\cite{Yang2024} & Optical + Infrared & $\Delta W1$, $\Delta W2 > 0.2$, $\mathrm{SN}_{W1}>5$, $\mathrm{SN}_{W2}>5$ and $\mathrm{SN}_{g}>4$\textsuperscript{c}\\
\hline
\textbf{Reference} & \textbf{Parameter}   & \textbf{Spectral Criteria}   \\
\hline
\citet{MacLeod2019} & Flux+S/N & $N_{\sigma}(\lambda)> 3$; $N_{\sigma}(\lambda) = (f_{bright} - f_{faint})$/ $\sqrt{\sigma^2_ {bright} + \sigma^2_{faint}}$\\
\cite{PottsVillforth2021}& Flux & $|\Delta f_{\lambda}|$ show multiple CLQ feature in H$\alpha$, H$\beta$ or continuum\textsuperscript{d} \\
\cite{Lopez2023}&Equivalent width (EW)&$> 3 \sigma$ change in the EW of broad H$\alpha$ and the H$\alpha$/[S ii] ratio\\
\cite{Yang2023}&Flux+S/N& same with \citet{MacLeod2019}\\
\cite{Zeltyn2024}&Flux+S/N&C (line) $\equiv$ F2/F1 - $\Delta$(F2/F1), max[C (line)] $>$ 2\\
\cite{Yang2024} & Flux & $f_{\mathrm{line,faint}}$/$f_{\mathrm{line,bright}} < 0.3$\\

\hline
\end{tabularx}
\begin{tablenotes}[flushleft]
\footnotesize
\item[$^a$] $W1$ and $W2$ are bands centered at wavelengths of 3.4, 4.6 $\mu \mathrm{m}$ observed by the Wide-field Infrared Survey Explorer (WISE). g is g-band magnitude from PS1, or DELS when available, or SDSS.
\item[$^b$] g and r are magnitudes from SDSS and PS1 measurements.
\item[$^c$] intrinsic variability SN=$\sigma_\mathrm{{band}}/\Delta{\sigma_{\mathrm{band}}}$, g is g-band magnitude from ZTF lightcurves.
\item[$^d$] The difference spectrum $|\Delta f_{\lambda}| = |f_{epoch 2} - f_{epoch 1}|$.
\end{tablenotes}
\end{threeparttable}
\end{table*}

Although the appearance or disappearance of broad emission lines is widely used to define CLAGNs, and various quantitative selection criteria have been proposed, visual inspection has remained the most commonly used as the final criterion in confirming CLAGN candidates.
In this work, we collect the CLAGNs from literatures that present the spectra in multiple phases. 
We perform a strict visual inspection for CLAGN identification. For sources where both H$\alpha$ and H$\beta$ are covered, a source is classified as a CLAGN only if both lines fully appear or disappear between the bright and faint states (SNR $>3\sigma$ at bright state and $<1\sigma$ at faint state), effectively excluding most intermediate-type AGNs (e.g., Type 1.5–1.9). Among the objects where only H$\beta$ is covered (H$\alpha$ is out of optical band), some might be Type 1.9s in their dim state. These sources are conservatively treated as Type 2 in our analysis.

Table \ref{Table:CLAGN} lists the number of CLAGNs from literature, and Table \ref{table: clagn} shows the details of all collected CLAGNs. We collect 228 CLAGNs in total, in which 160 of them have SMBH mass, bolometric luminosity and Eddington ratio measurements from \citet{Shen2022,Yang2024}. Among them, 152 sources have redshift $z < 0.8$ to ensure the $\mathrm{H}\alpha$ or $\mathrm{H}\beta$ observations in the optical spectra.

\begin{table}[h!]
\centering
\caption{Number of CLAGNs from literature.}\label{Table:CLAGN}
\begin{tabular}{ccc}

\hline
\textbf{Reference} & \textbf{Number}   \\
\hline
\cite{Zeltyn2024}& 72 \\
\cite{Yang2024}&70\\
\cite{Yang2018}& 26\\
\cite{MacLeod2019}&17\\
\cite{MacLeod2016}&10\\
\cite{Lopez2023}&9\\
\cite{Sheng2019}&6\\
\cite{Frederick2019}&6\\
\cite{Lopez2022}&4\\
\cite{PottsVillforth2021}&4\\
\cite{Yang2023}&4\\
\hline
\end{tabular}

\end{table}

\subsection{Control Sample}

\begin{figure*}[htbp]
    \centering
    \includegraphics[width=\textwidth]{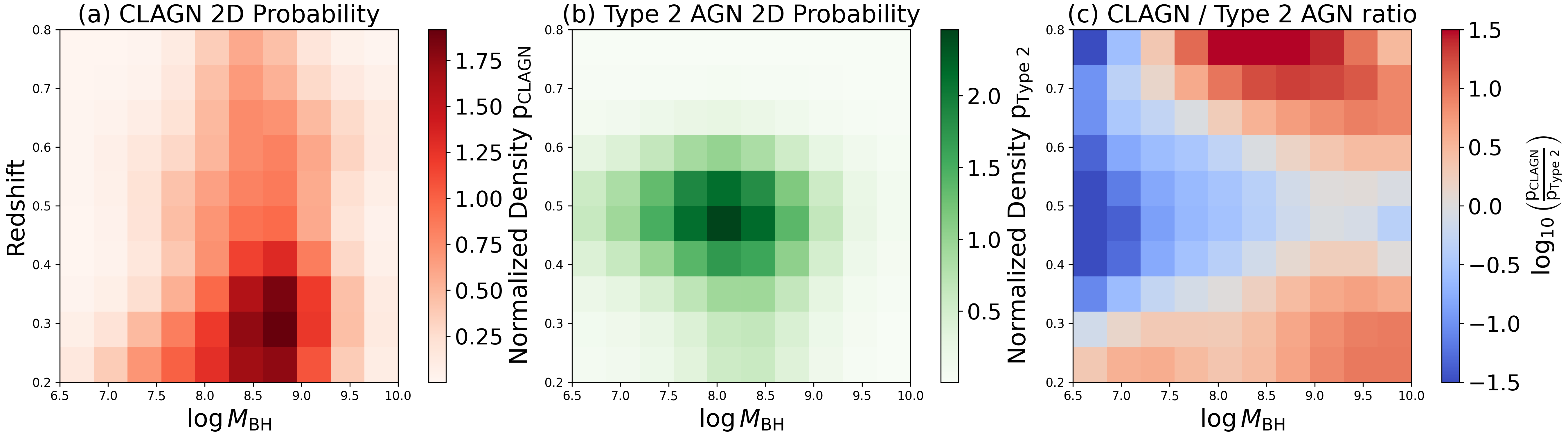}
    \caption{
        Two-dimensional distributions of CLAGNs (a) and Type 2 AGNs (b) in the $\log M_{\rm BH}$–$z$ plane, and their logarithmic ratio (panel c). These distributions are used to construct a control sample of Type 2 AGNs matched to the CLAGNs in both dimensions via the probability-based sampling. Type 2 sample exhibits lower source density near all four corners of the 2D parameter space, resulting in larger sampling uncertainties and stronger deviations in the ratio. 
    }
    \label{fig:clagn_type2_ratio}
\end{figure*}

We construct a control sample consisting of Type 1 AGNs, Type 2 AGNs, and extremely variable quasars (EVQs) with the same physical parameters including redshift and SMBH mass with our CLAGNs to compare the photometric variability between CLAGNs and the control samples. For all kinds of control sample, we limit the redshift $z \textless 0.8$, to ensure spectroscopic coverage in the observed $\mathrm{H}\beta$ or $\mathrm{H}\alpha$ wavelength windows.

EVQs are a population of sources that exhibit large optical photometric variability, with CLAGNs potentially comprising a subsample \citep{Ren2022}. These EVQs were selected by the criterion: $|\Delta g_{\mathrm{max}}|>1$ mag and $|\Delta r_{\mathrm{max}}|>0.8$ mag where $|\Delta g_{\mathrm{max}}|$ and $|\Delta r_{\mathrm{max}}|$ represent the maximum differences in $g$- and $r$-band magnitudes between any two epochs in the light curves. \citet{MacLeod2019} adopted a similar selection to identify CLAGN candidates and 
constructed the follow-up spectroscopic confirmation. 
About $20\%$ of EVQs were classified as CLAGNs, while the others, though highly variable, did not exhibit broad-line changes large enough to satisfy the disappearance/emergence criterion \citep{MacLeod2019}.
This makes EVQs a suitable comparison sample for CLAGNs, under the assumption that the difference arises from the presence or absence of spectral transitions. Using the EVQs from \citet{Ren2022}, and since we are interested in accretion- rather than jet-related variability, we cross-match the FIRST survey in radio band \citep{Becker1995} and rule out 669 radio-detected objects.

\begin{figure*}[htb!]
\epsscale{1}
\plotone{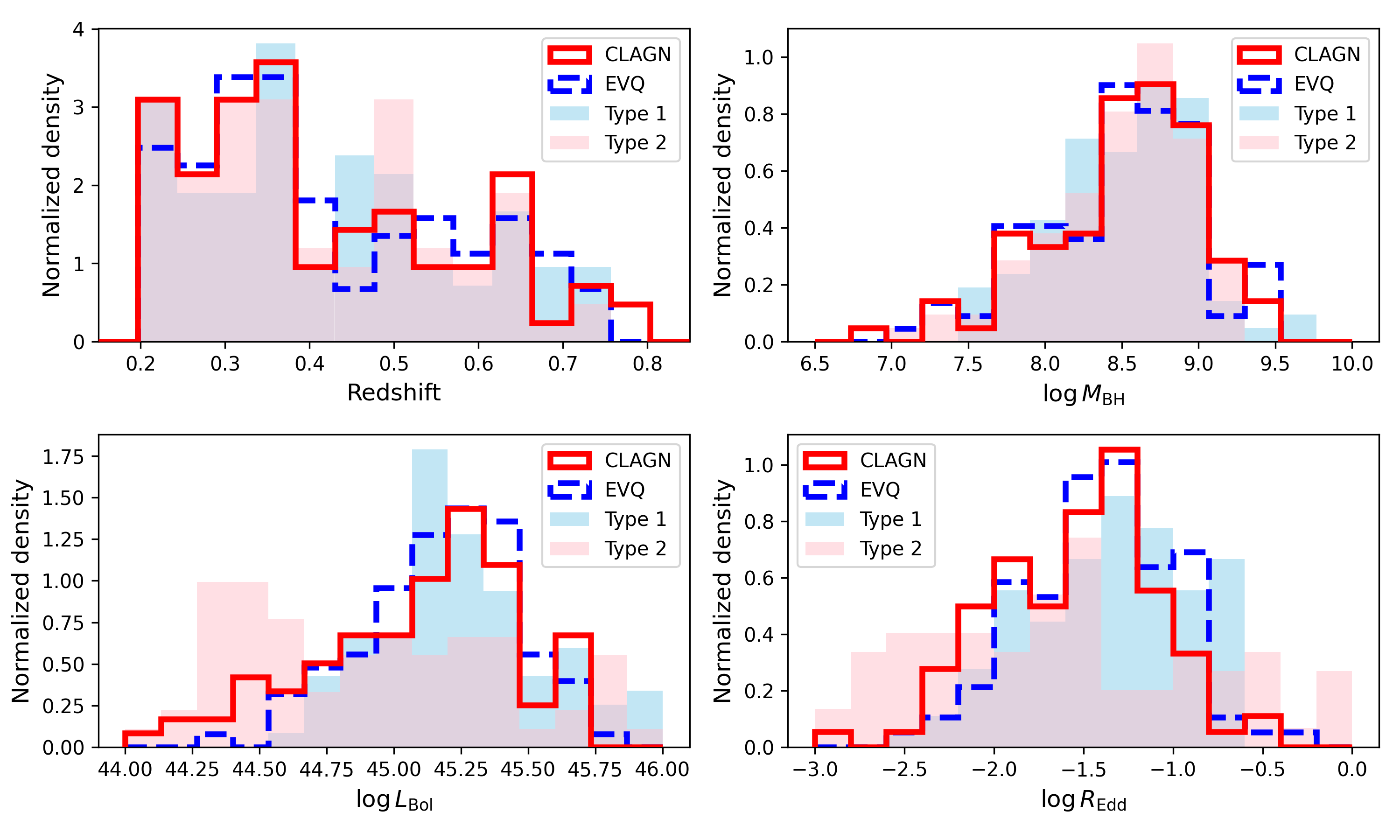}
\caption{{AGN properties for the CLAGN and control samples obtained from the SDSS catalog \citep{Shen2022}. The skyblue and pink regions refer to Type 1 and Type 2 AGN samples respectively. The blue dashed line represents the distribution of EVQ samples and the red solid line shows the distribution of CLAGN samples.}}
\label{fig:AGN_property}
\end{figure*}

For Type 1 AGNs, we start from the SDSS DR 16 Quasar Catalog \citep{Lyke2020}, which contains 750,414 quasars from SDSS I, II, III, and IV. We select the sources with full-width-at-half-maxima (FWHM) of $\mathrm{H\beta}$ profile, $\mathrm{FWHM_{H\beta} > 1500km\, s^{-1}}$. 

Type 2 AGNs were selected from the Million Quasar Catalog (MQC) v7.2 \citep{flesch2021millionquasarsmilliquasv72}. We down-selected sources with type/class ``K'' (narrow-line Seyferts) and/or "N" (narrow-line Quasi-Stellar Objects; QSOs) and crossmatched with the Portsmouth Stellar Kinematics and Emission Line Fluxes Value Added Catalog from SDSS \citep{Thomas2013}. Our final sample consists of 152 CLAGNs, 3690 EVQs, 73963 Type 1 AGNs and 5981 Type 2 AGNs. 

We also conduct a control sample of CLAGNs in on/off-state. We select the sources identified in 2015-2023 as CLAGN, and confirm their ZTF lightcurve in a stable phase, using the last spectral classification to distinguish on-state and off-state. Figure \ref{fig:clagn_lightcurves} shows three examples of cleaned, binned light curves, for the turn-on stable, turn-on unstable and turn-off stable AGNs, respectively.
The sources are recently unstable in ZTF light curves are excluded in this control sample. There are 48 off-state light curves and 80 on-state light curves in total.

Given that redshift and black hole mass are less model-dependent in measurement and significantly influence quasar behavior \citep{Hopkins2007,Kelly2009}, matching these parameters is crucial. 
For CLAGN, EVQ and Type 1 AGN, the black hole mass, $M_{\mathrm{BH}}$, is derived using the single-epoch virial estimators \citep{Vestergaard2006}, which rely on the broad emission line width and continuum luminosity as proxies for the virial motion of the broad-line region (BLR). The bolometric luminosity ($L_{\mathrm{bol}}$) is estimated based on the continuum luminosity, with a preference for the 3000$\mathrm{\AA}$ luminosity when available, as it is less affected by host galaxy contamination than the 5100$\mathrm{\AA}$ luminosity and exhibits lower sensitivity to reddening and variability compared to the 1350$\mathrm{\AA}$ luminosity \citep{Shen2022}. The Eddington ratio is computed as $R_{\mathrm{Edd}} \equiv L_{\mathrm{bol}}/L_{\mathrm{Edd}}$, where the Eddington luminosity is given by  
$L_{\mathrm{Edd}} = 1.3 \times 10^{38} \left(\frac{M_{\mathrm{BH}}}{M_{\odot}}\right) \, \mathrm{erg\,s}^{-1}.$ 
For EVQ and Type 1 AGN, these quantities are adopted from \citet{Shen2022}.

The black hole mass of Type 2 AGN is estimated using the $M$–$\sigma$ relation from \citet{Ferrarese2002}:
\begin{equation}
    M_{\mathrm{BH}}=(1.66 \pm 0.32) \times 10^8 M_{\odot}\left(\frac{\sigma_*}{200 \mathrm{~km} \mathrm{~s}^{-1}}\right)^{4.58 \pm 0.52},
\end{equation}
where the stellar velocity dispersion $\sigma_*$ is taken from the Portsmouth group galaxy properties catalog \citep{Thomas2013}.

We adopt the extinction-corrected [O III] $\lambda5007$ line fluxes from the Portsmouth catalog \citep{Thomas2013}, which are corrected for the diffuse dust component affecting the overall spectral shape by using the \citet{Calzetti2001} obscuration curve and not including Balmer-decrement based dust corrections. We then use them to estimate the bolometric luminosity, $L_{\mathrm{bol}}$, following \citet{Pennell2017}, as
\begin{equation}
    L_{[\mathrm{O\,III}]} = 4\pi D_L^2 F_{[\mathrm{O\,III}]}, 
\end{equation}
\begin{equation}
    \log \left(\frac{\mathrm{L}_{\text {bol }}}{\mathrm{erg~s}^{-1}}\right)=\log \left(\frac{\mathrm{L}_{\left[\mathrm{O\,III}\right]}}{\mathrm{erg~s}^{-1}}\right)+(3.532 \pm 0.059),
\end{equation}
where $D_L$ is the luminosity distance derived from the redshift.

We note that this approach is subject to systematic uncertainties, including the intrinsic scatter in the $\mathrm{M}$–$\sigma$ relation and the potential contamination of [O\,\textsc{iii}] luminosity by star formation or shock excitation.

To construct a control sample of EVQs, Type 1 AGNs, and Type 2 AGNs that matches the CLAGNs in both black hole mass and redshift, we adopt a two-dimensional probability matching scheme, which statistically reproduces the distributions of CLAGNs in $z$ and $M_{\rm BH}$. To maintain consistency across samples, we restricted all sources to $z>0.2$, as the EVQ sample contains very few objects at redshifts lower than 0.2. After this cut, 90 CLAGNs remained in our sample.
Specifically, the black hole mass range is $10^{6.5}$ to $10^{10} M_\odot$ and is divided into 10 logarithmic bins, while the redshift range is $0.2<z<0.8$ and is divided into 10 uniform bins for the probability matching. The left and middle panels of Figure  \ref{fig:clagn_type2_ratio}  show the 2D distributions of CLAGNs and Type 2 AGNs in the $(\log M_{\rm BH}, z)$ plane, while the right panel presents the logarithmic ratio between the two. 
Each object in the control sample was assigned a weight proportional to the ratio of these densities, and 90 sources were randomly drawn using the normalized weights as selection probabilities. 
As seen in the figure, the Type 2 AGN sample has fewer sources in the four corners of the 2D parameter space, which leads to higher statistical uncertainty when sampling control sources from these regions. As a result, even though the matching procedure is probabilistic, the reconstructed control sample may show larger fluctuations at the distribution edges, particularly where the original Type 2 density is low. This illustrates a limitation of the matching method in regions where the control sample has intrinsically sparse coverage. In our analysis, we also conducted a comparison without controlling for redshift and black hole mass, by drawing random subsamples of equal size from the full control sets, and the results did not show significant differences. 

To ensure fair comparisons, we adopted a 1:1:1:1 matching scheme so that the scatter in their distributions is comparable.
After applying the matching, we obtained final samples of 90 CLAGNs, 90 EVQs, 90 Type 1 AGNs, and 90 Type 2 AGNs. The resulting control samples exhibit $M_{\mathrm{BH}}$ and z distributions statistically consistent with those of the CLAGNs, as confirmed by the Kolmogorov-Smirnov (KS) test.
Figure \ref{fig:AGN_property} shows the normalized density of bolometric luminosity, $L_{\mathrm{bol}}$ and Eddington ratio, $R_{\mathrm{Edd}}$. 
Values of $L_{\mathrm{bol}}$ and $R_{\mathrm{Edd}}$ are adopted from \citet{Shen2022} when available, so the distribution shown is mostly for the bright-state spectra. For the sources in \citet{Yang2024}, both dim- and bright-state measurements are available, and we adopt the bright-state values to remain consistent with \citet{Shen2022}.

\section{Results} \label{sec:result}
\subsection{Color variability}

Previous studies have shown that CLAGNs generally follow the well-known ``bluer-when-brighter" trend in the optical bands, similar to normal Type 1 AGNs \citep{Yang2018}. To investigate this in more detail, we analyzed the color variability of CLAGNs using ZTF light curves in the $g$- and $r$- bands. The light curves were binned in 3-day intervals, and for epochs with observations in both bands, we computed the $g-r$ (hereafter we denote $g$ and $r$ as the ZTF $zg-$ and $zr$-band) color and the change in $zg$-band magnitude, $\Delta g$. Here, a negative $\Delta g$ indicates that the object becomes brighter, and a negative color variability value corresponds to a bluer color.

\begin{figure}[htb!]
\epsscale{1}
\plotone{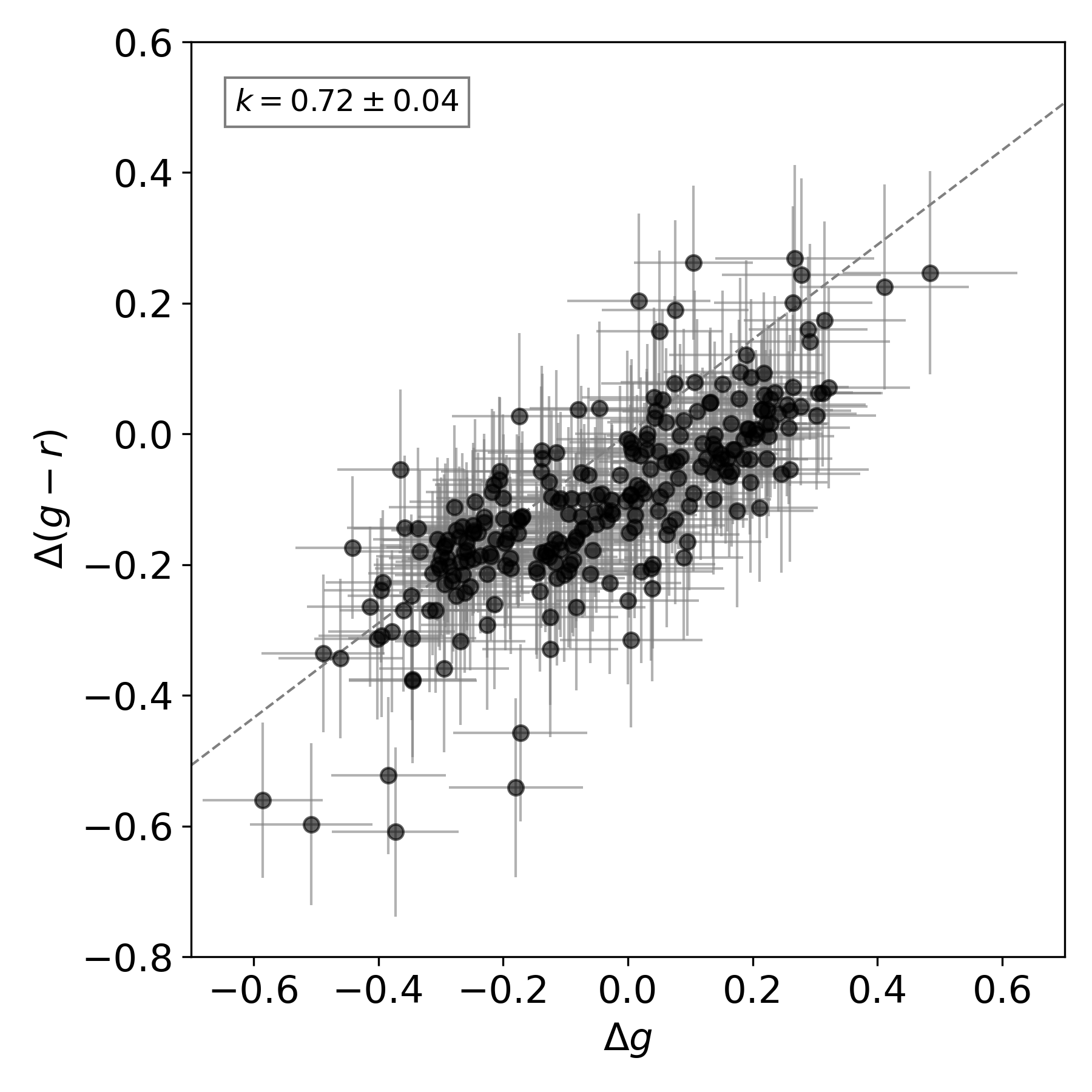}
\caption{{Color variability vs. magnitude variability for the CLAGN, SDSS J0040+1609, in the optical band. The gray dashed line represents the best-fit linear relation. The data confirm a bluer-when-brighter trend. }}
\label{fig: color_single}
\end{figure}

Using this dataset, we constructed the color-magnitude diagrams ($zg-zr$ vs. $zg$) for each CLAGN, with representative examples SDSS J0040+1609 shown in Figure \ref{fig: color_single}. To quantify the color variability, we fit the color-magnitude relation with a linear function:

\begin{equation}
        \Delta (g-r)=k\times\Delta g,
\end{equation}

\noindent where $k$ represents the slope of the relation. A positive $k$ value indicates a ``bluer-when-brighter'' trend, while a negative $k$ value corresponds to the ``redder-when-brighter" pattern.

\begin{figure}[htb!]
\epsscale{1}
\plotone{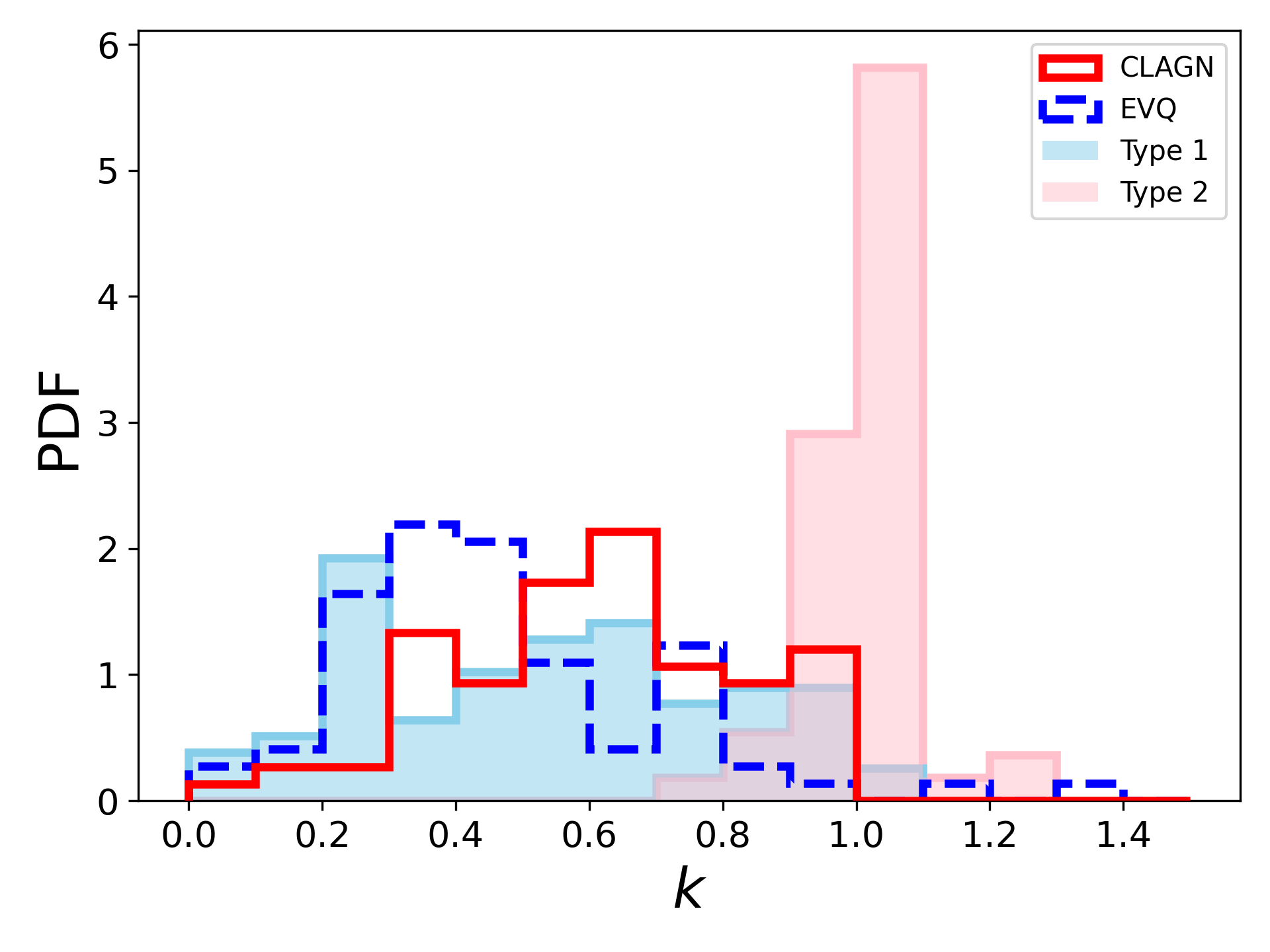}
\caption{{Distributions of the $k$ values (slopes) from the linear fits to the color-magnitude variation data points for CLAGNs and the corresponding control samples. The lines have the same meaning as in Figure 1. Type 1 AGN and Type 2 AGN share different distribution while CLAGN located in the middle region.}}
\label{fig: CMhist}
\end{figure}

To ensure the reliability of the $k$ values, we computed the Pearson correlation coefficient during the fitting procedure and excluded sources with a correlation significance level of $p > 0.05$. Figure \ref{fig: CMhist} displays the distribution of $k$ values for CLAGNs and their corresponding control samples. Type 1 AGNs exhibit a central $k$ value around 0.50, consistent with the typical variability behavior of unobscured AGNs. Type 2 AGNs, in contrast, show a steeper slope, with $k$ values centered around 1.0. The distribution of EVQs is similar to Type 1 AGNs, while CLAGNs occupy an intermediate region between Type 1 and Type 2 AGNs. This suggests that CLAGNs exhibit color variability properties distinct from both normal Type 1 and Type 2 AGNs, potentially reflecting their transitional nature.

We illustrate the differences in color variability among different samples and summarize the mean and standard deviation of the color variability slopes, $k$, in Table \ref{tab:color_variability}. The values confirm the observed trend in Figure \ref{fig: CMhist}, with Type 1 AGNs showing the smallest slopes and Type 2 AGNs exhibiting the steepest changes. EVQs show little difference with Type 1 AGNs while CLAGNs fall in between, further supporting the intermediate nature.

\begin{figure}[htb!]
\epsscale{1}
\plotone{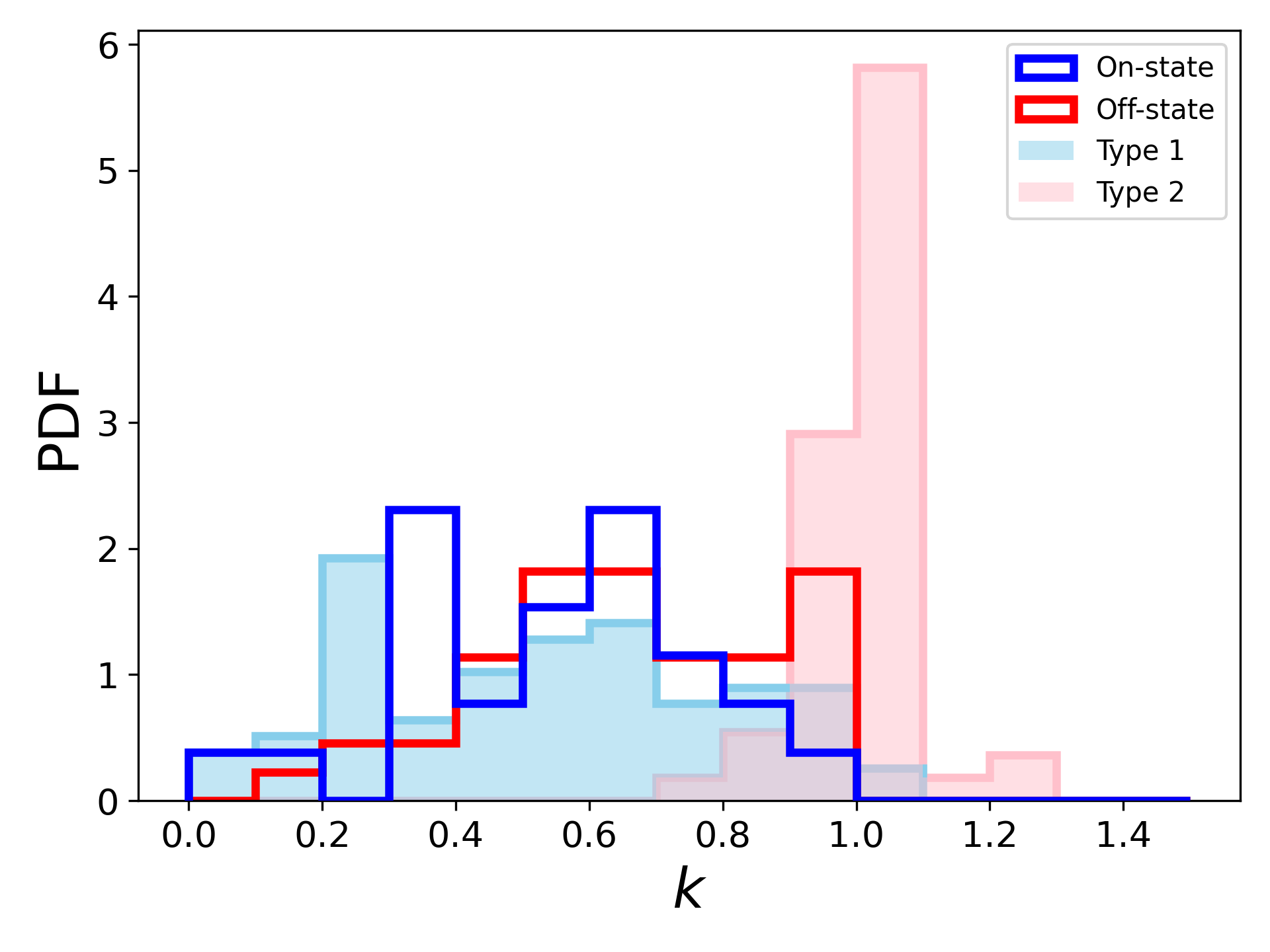}
\caption{{Distributions of the $k$ values (slopes) from the linear fits to the color-magnitude variation data points for on/off-state and the corresponding control sample. The skyblue and pink lines represent Type 1 and Type 2 AGNs. The blue and red lines show the on-state and off-state sources, which have a similar distribution located in the middle region.}}
\label{fig: CMhist_onoff}
\end{figure}

To further investigate the behavior of CLAGNs in different spectral states, we separately analyzed the color-magnitude slopes for the on-state and off-state epochs (Figure \ref{fig: CMhist_onoff}). Notably, the $k$ distribution for the off-state CLAGNs does not align with that of normal Type 2 AGNs, and similarly, the on-state distribution differs from that of normal Type 1 AGNs. To further quantify the similarity between the color variability slopes of on-state / off-state CLAGNs and those of the control samples, we performed KS tests on these samples. Table \ref{tab:ks_test} represents the KS statistics and p-values for different comparisons. 

The KS test results indicate that both on- and off-state CLAGNs are generally more similar to the overall CLAGN sample than to their respective control AGN types. Specifically, on-state CLAGNs are closer to the overall CLAGN distribution than to Type 1 AGNs ($p = 0.631$ vs. $p = 0.363$), and off-state CLAGNs are closer to the overall CLAGN distribution than to Type 2 AGNs ($p = 0.848$ vs. $p = 6.96\times10^{-18}$).
These findings support the idea that CLAGN variability is not simply a direct transition between Type 1 and Type 2, thus the variability properties of CLAGNs are not simply a combination of distinct Type 1 and Type 2 behaviors, but represent an intermediate mode.

\begin{table*}[htb!]
\centering
\caption{Mean and standard deviation of color variability slopes, $k$, and variability metrics for different samples (updated).}
\label{tab:color_variability}
\begin{tabular}{lcccc}
\hline
Sample & $k$ & $\mathrm{log(1+\sigma_{QSO})}$ & $\mathrm{log(1+\sigma_{var})}$ & $f(\sigma_{\mathrm{QSO}}>3)$ \\
\hline
CLAGN & $0.61 \pm 0.21$ & $1.025 \pm 0.324$ & $1.440 \pm 0.429$ & 0.873 \\
On-state & $0.55 \pm 0.22$ & $1.063 \pm 0.282$ & $1.568 \pm 0.383$ & 0.926 \\
Off-state & $0.65 \pm 0.21$ & $0.987 \pm 0.344$ & $1.419 \pm 0.454$ & 0.830 \\
EVQ & $0.46 \pm 0.24$ & $1.031 \pm 0.351$ & $1.673 \pm 0.493$ & 0.881 \\
Type 1 AGN & $0.53 \pm 0.27$ & $0.923 \pm 0.352$ & $1.421 \pm 0.500$ & 0.874 \\
Type 2 AGN & $1.02 \pm 0.08$ & $0.281 \pm 0.137$ & $0.579 \pm 0.280$ & 0.169 \\
\hline
\end{tabular}
\end{table*}

\begin{table*}[htb!]
    \centering
    \caption{Kolmogorov-Smirnov test results for the on/off-state CLAGNs and control samples (updated)}
    \label{tab:ks_test}
    \begin{tabular}{lcccccc}
        \hline
        \hline
        Comparison & $D_k$ & $p_k$ & $D_{\log(1+\sigma_{\rm QSO})}$ & $p_{\log(1+\sigma_{\rm QSO})}$ & $D_{\log(1+\sigma_{\rm var})}$ & $p_{\log(1+\sigma_{\rm var})}$ \\
        \hline
        On-State vs. Type 1 & 0.2051 & 3.63$\times10^{-1}$ & 0.2095 & 2.80$\times10^{-1}$ & 0.2312 & 1.86$\times10^{-1}$ \\
        Off-State vs. Type 2 & 0.8364 & 6.96$\times10^{-18}$ & 0.7080 & 9.70$\times10^{-14}$ & 0.6890 & 5.75$\times10^{-13}$ \\
        On-State vs. CLAGN & 0.1610 & 6.31$\times10^{-1}$ & 0.1303 & 8.31$\times10^{-1}$ & 0.1430 & 7.42$\times10^{-1}$ \\
        Off-State vs. CLAGN & 0.1094 & 8.48$\times10^{-1}$ & 0.1045 & 8.60$\times10^{-1}$ & 0.1131 & 7.92$\times10^{-1}$ \\
        On-State vs. EVQ & 0.3590 & 9.84$\times10^{-3}$ & 0.2183 & 2.43$\times10^{-1}$ & 0.2222 & 2.25$\times10^{-1}$ \\
        Off-State vs. EVQ & 0.4848 & 1.86$\times10^{-6}$ & 0.2302 & 6.72$\times10^{-2}$ & 0.2191 & 9.21$\times10^{-2}$ \\
        \hline
    \end{tabular}
\end{table*}

\subsection{Structure Function (SF)}
The structure function (SF) is a widely used statistical tool for characterizing the variability of AGNs by quantifying how the amplitude of variability depends on the time separation between observations. For objects with $n_{\Delta t}$ light curve data points, the observed variability amplitude between epochs with separation $\Delta t$ is \citep{Kozlowski2016},
\begin{equation}
    SF_{ij,obs}=\sqrt{\frac{1}{n_{\Delta t}} \sum_{i,j}\left\{\left[m_i-m_j\right]^2-\sigma^2_i-\sigma^2_j\right\}}.
\end{equation}
Here, the summation is taken over all epoch pairs $i,j$ with time separations that fall within a given time lag bin $\Delta t$ in the rest frame. The measured magnitudes at epochs $i$ and $j$ are denoted by $m_i$ and $m_j$, and their respective photometric uncertainties are $\sigma_i$ and $\sigma_j$ with a 0.9 factor \citep{Kim2024}.

This structure function can be described with a power-law model \citep{Schmidt2010}:
\begin{equation}
    SF_{i j, \text { model }}=A\left(\frac{\left|t_i-t_j\right|}{1 \mathrm{yr}}\right)^\gamma ,
\end{equation}
where $A$ refer to the typical variability for 1 year, and $\gamma$ is the logarithmic gradient of this mean change in magnitude. 

We use a simple Markov chain Monte Carlo (MCMC) approach to estimate the parameters $A$ and $\gamma$. The likelihood function is given by:

\begin{equation}
    \mathcal{L}(A, \gamma)=\prod_{i, j} L_{i, j},
\end{equation}
assuming the magnitude difference in CLAGN is independent. The $L_{i,j}$ is the likelihood of observing magnitude difference $\Delta m_{i,j}$ between two light curve points separated by $\Delta t_{i,j}$ for a single source. 

We assume an underlying Gaussian distribution of $\Delta m$ values and Gaussian photometric errors, then the likelihood can be defined as
\begin{align}
\begin{split}
    L_{i, j}=&\prod_{i,j} \frac{1}{\sqrt{2 \pi \sigma_{S F,ij}^2}} \\
    &\times\exp \left(-\frac{\left(S F_{i j}-S F_{i j, \text { model }}\right)^2}{2 \sigma_{S F,ij}^2}\right) .
\end{split}
\end{align}

The uncertainty in the structure function, $\sigma_{SF,ij}$, was estimated using a bootstrap resampling method. For each time bin, we randomly selected 80\% of the available data points, recalculated $SF_{ij}$ 1000 times, and took the standard deviation of these realizations as $\sigma_{SF,ij}$.

For the prior distributions, we adopted an uninformative log-uniform prior for $A$, $P(A) \propto 1/A$ with $A \in (0,1)$, and an arctangent-uniform prior for $\gamma$, $P(\gamma) \propto 1/(1+\gamma^2)$, $\gamma \in (0,+\infty)$ \citep{Schmidt2010}. 

To ensure the robust estimates, we used logarithmic binning in the time domain to balance the number of data points in each bin.  
We cut out the vibration tail for each source\citep{Tang2023}. For each light curve, the cut fraction was determined by scanning within the 0.2–0.8 range and selecting the fraction that yields the best linear fit of the structure function. Sources with fewer than 20 data points remaining after this cut were excluded. The parameters $A$ and $\gamma$ were then estimated for the remaining CLAGNs and control sample objects.

\begin{figure}[htb!]
\epsscale{1}
\plotone{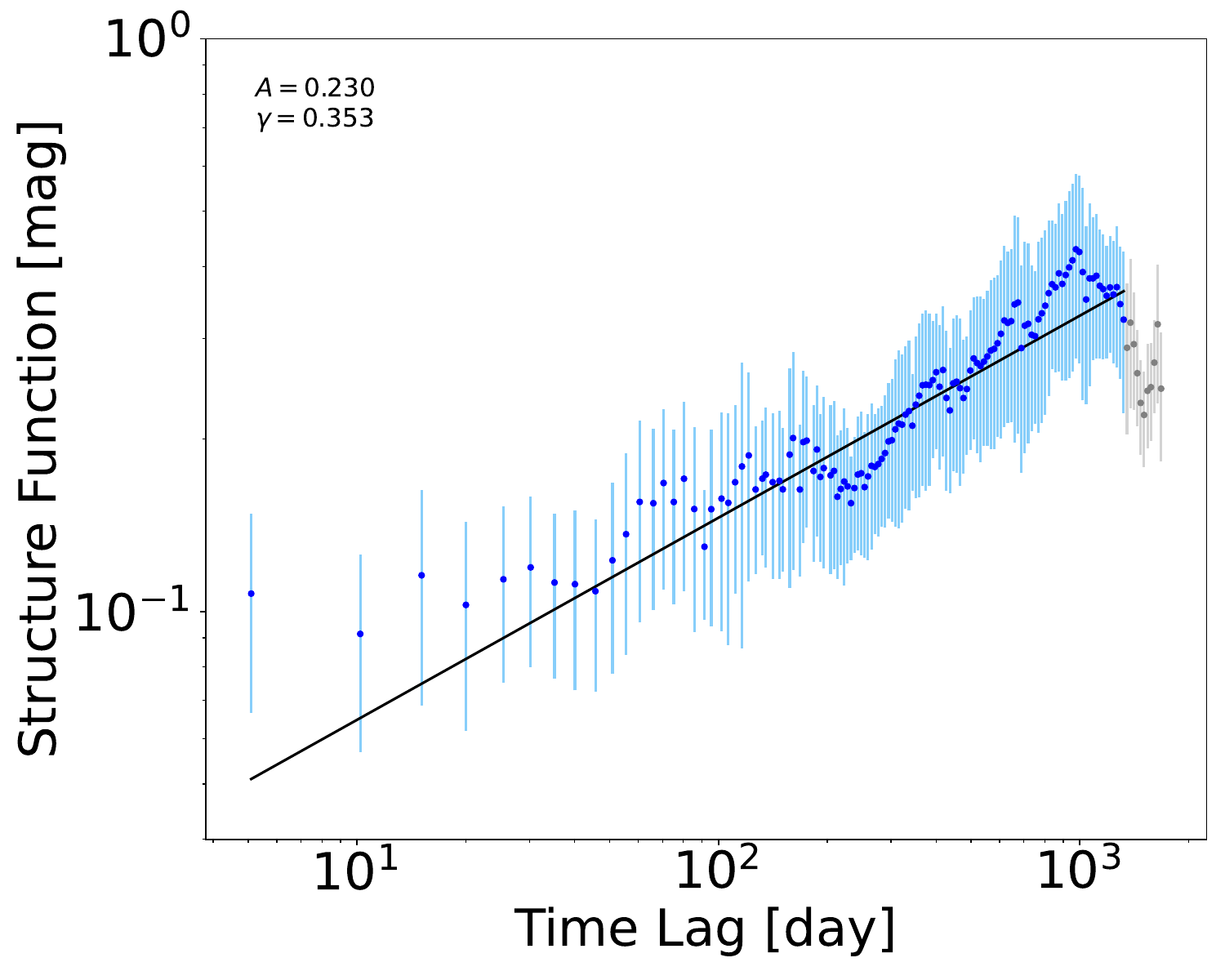}
\caption{{Structure function of a CLAGN SDSS J0040+1609. The points represent the observed SF, while the black line shows the best-fit power-law model, characterized by the estimated parameters $A$ and $\gamma$. The gray points were cut off to avoid the noisy vibration tail at longer time lags.}}
\label{fig: SF_sample}
\end{figure}

There are 71 CLAGNs, 66 EVQs, 75 Type 1s, and 38 Type 2s considered in this section. 
Most of the excluded sources are Type 2s at slightly higher redshifts, but the overall redshift and black hole mass distributions are not significantly affected, and thus the impact on our results is negligible.
Figure \ref{fig: SF_sample} illustrates the procedure for an example of a CLAGN.

\begin{figure}[htb!]
\epsscale{1}
\plotone{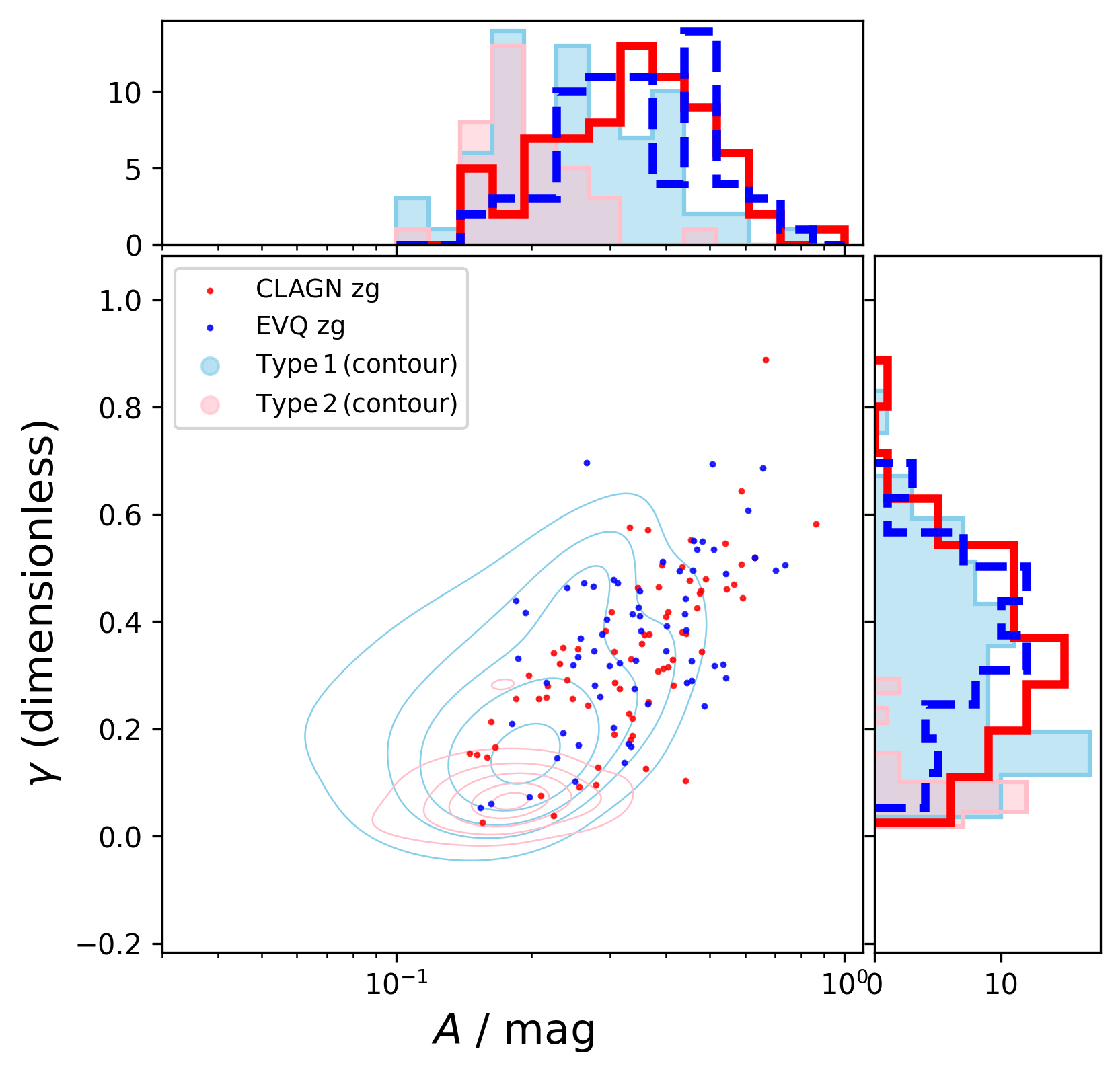}
\caption{{Comparison of $A$ and $\gamma$ for different AGN samples: CLAGN in red points, EVQ in blue points, Type 1 AGNs in skyblue contour, and Type 2 AGNs in pink contour. The top and right panel show the histogram for $A$ and $\gamma$. The scatter plot highlights distinct clustering: CLAGNs and EVQs generally exhibit larger $A$ values compared to Type 1 AGNs and Type 2 AGNs, indicating greater variability. EVQs and Type 1 AGNs tend to have higher $\gamma$ than CLAGN, while Type 2 AGNs are concentrated at even lower values. This demonstrates significant differences in variability properties across AGN types.}}
\label{fig: SF}
\end{figure}

Figure \ref{fig: SF} presents the distribution of the variability characteristics, quantified by the best-fit $A$ and $\gamma$ for the selected sources with $zg$ band data. The histograms along the axes of the two-dimensional scatter plot show the parameter projected distribution of the CLAGNs and control samples. Notably, our results, particularly the $A$–$\gamma$ distribution of the Type 1 AGNs, are broadly consistent with those of QSOs reported by \citet{Schmidt2010}, although our parameter range appears narrower, likely due to the smaller sample size.

Overall, these results highlight that CLAGNs exhibit stronger and more rapidly evolving variability compared to typical Type 2 AGNs, with a distribution of SF parameters more similar to EVQs and Type 1 AGNs. The distinction in the structure function properties between CLAGNs and other AGN types provides further insight into their unique variability behavior.

\subsection{Variability metric}
Type 1 and Type 2 AGNs exhibit distinct variability properties due to differences in their accretion physics and dust obscuration. While Type 1 AGNs show strong, stochastic variability, primarily driven by fluctuations in the accretion disk, Type 2 AGNs exhibit significantly weaker variability, likely due to the presence of a dusty torus that obscures the variable continuum \citep{Kovacevic2025}. Since CLAGNs experience transition between these two types, it is crucial to investigate whether their variability characteristics also evolve during different phases of the transition, as this can provide insight into the underlying physical mechanisms governing CLAGN behavior.

\begin{figure*}[htb!]
\epsscale{1}
\plotone{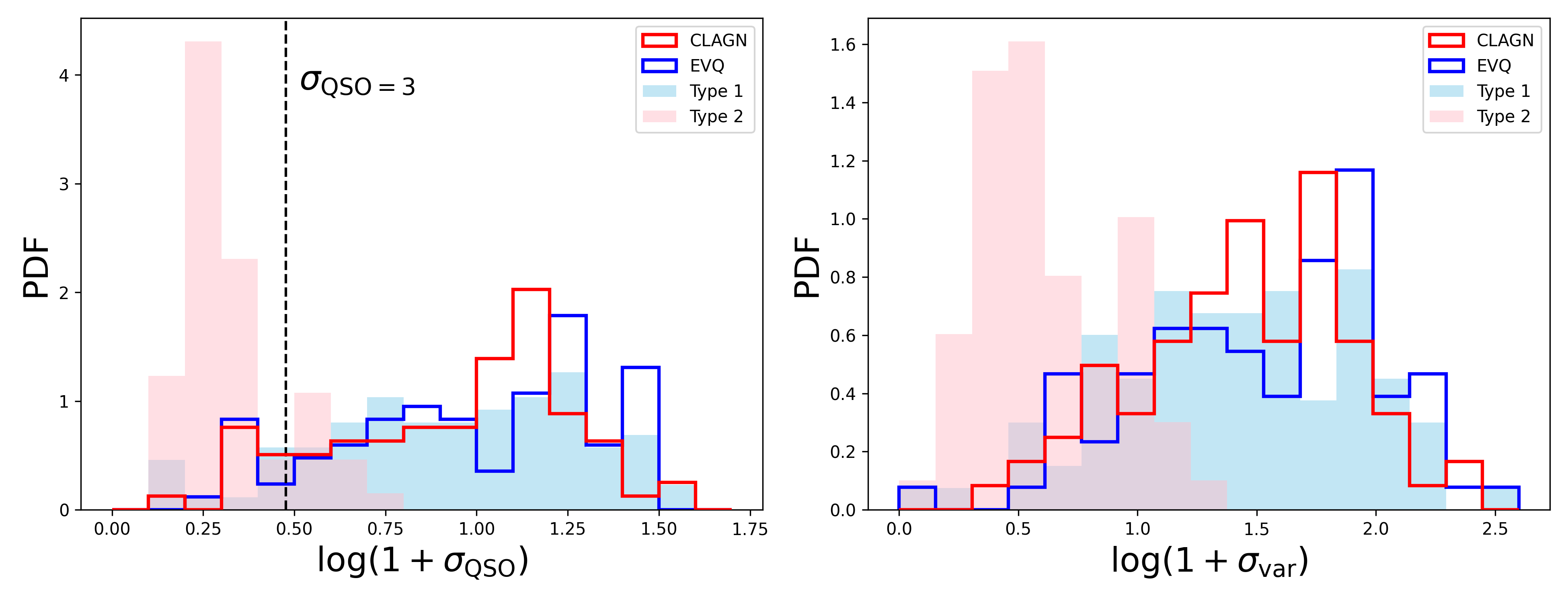}
\caption{{Comparison of $\sigma_{\mathrm{QSO}}$ and $\sigma_{\mathrm{var}}$ for different AGN samples: CLAGN in red line, EVQ in blue line, Type 1 AGNs in skyblue region, and Type 2 AGNs in pink region. The majority of Type 1 and 2 AGNs are well separated while CLAGNs and EVQs locate in the middle region.}}
\label{fig: metric}
\end{figure*}

\begin{figure*}[htb!]
\epsscale{1}
\plotone{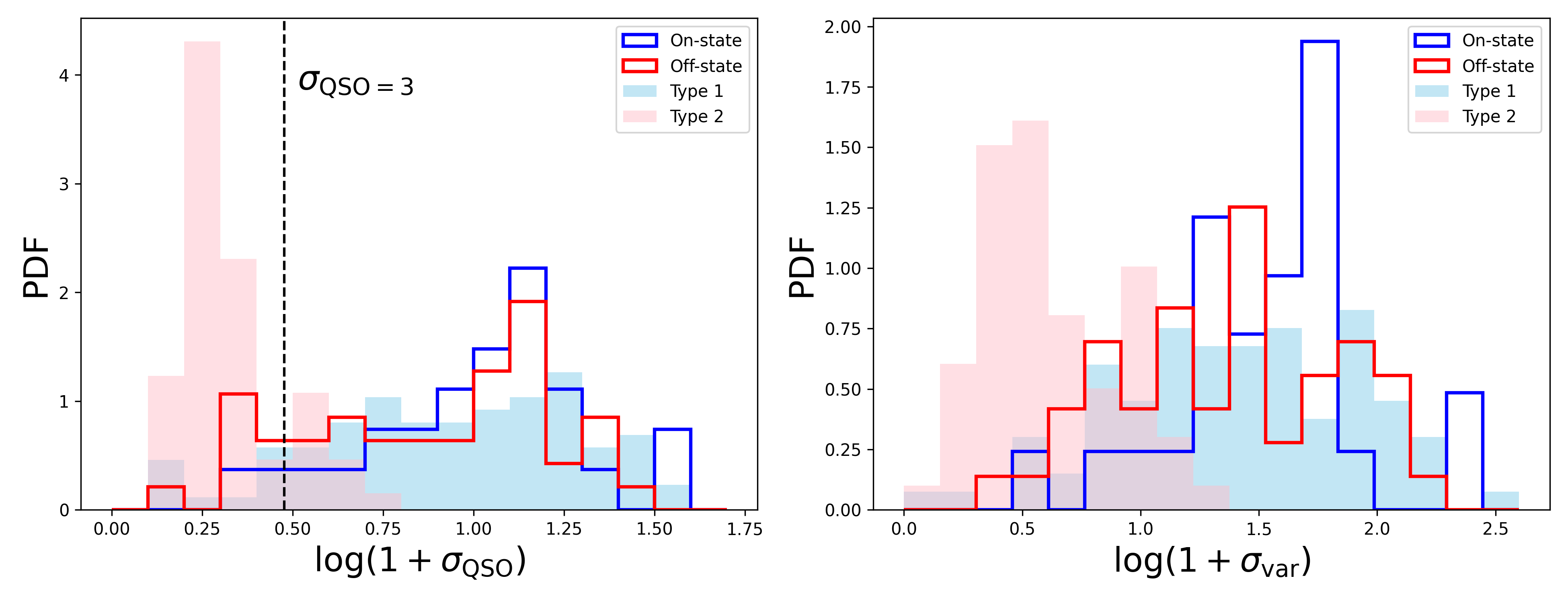}
\caption{{Comparison of $\sigma_{\mathrm{QSO}}$ and $\sigma_{\mathrm{var}}$ for different AGN samples: off-state in red line, on-state in blue line, Type 1 AGNs in skyblue region, and Type 2 AGNs in pink region. Both on-state and off-state CLAGNs locate in the middle region.}}
\label{fig: metric_on}
\end{figure*}

In this section, we quantify the variability statistics using the \texttt{$\mathrm{qso\_fit}$} software\footnote{\text{https://github.com/wt-liao/quasar\_drw}} \citep{Butler2011}, which was developed to separate (type 1) quasars from other point-like sources (e.g., stars) in SDSS Stripe 82 based on their $g$ band light curves. 

We adopt two parameters in the variability metric to describe the variability. (i) $\sigma_{\mathrm{var}}$ assesses whether the source is variable, and the significance of variability; (ii) $\sigma_{\mathrm{QSO}}$ quantifies the source variability prefer to be described by a DRW model rather than a time-independent variable Gaussian signals. The details and definition are presented in Appendix \ref{Sec: Appendix}. 
Combined, these metrics have been proven to efficiently separate quasars from stars with high completeness ($>$99\%) and low contamination ($<$3\%) in SDSS Stripe 82 \citep{Butler2011}. In particular, $\sigma_{\mathrm{QSO}} > 3$ has been used to be a indicator of quasar-like variability, and thus serves as an effective selection threshold. Accordingly, in addition to computing both variability metric parameters, we also evaluate the fraction of sources with $\sigma_{\mathrm{QSO}} > 3$, $f(\sigma_{\mathrm{QSO}}>3)$, in each sample for further comparison.

As shown in Figure \ref{fig: metric}, Type 1 and Type 2 AGNs exhibit well-separated distributions in both $\sigma_{\mathrm{QSO}}$ and $\sigma_{\mathrm{var}}$. Type 1 AGNs display high values of both parameters, reflecting their strong and DRW-like variability, which is characteristic of accretion disk fluctuations. In contrast, Type 2 AGNs have lower values of $\sigma_{\mathrm{QSO}}$ and $\sigma_{\mathrm{var}}$, indicating significantly weaker variability, likely due to the obscuration of the accretion disk by the dusty torus.

Rather than exhibiting a bimodal or double-peak distribution, CLAGNs and EVQs occupy a continuous distribution centered between Type 1 and Type 2 AGNs, or more close to Type 1 AGNs. Meanwhile, their $f(\sigma_{\mathrm{QSO}}>3)$ values are more close to Type 1 AGNs. 

Similar to the color-magnitude relation, we compare the distributions of on-state and off-state CLAGNs with those of Type 1 and Type 2 AGNs in the $\log(1 + \sigma_{\mathrm{QSO}})$ vs. $\log(1 + \sigma_{\mathrm{var}})$ plane (g band) in Figure \ref{fig: metric_on}. Notably, for $\sigma_{\mathrm{QSO}}$, the on-state CLAGNs do not exhibit a strong resemblance to typical Type 1 AGNs, nor do the off-state CLAGNs show a clear similarity to Type 2 AGNs. Instead, both states tend to occupy regions closer to the overall CLAGN distribution, suggesting that the variability properties of CLAGNs are not simply a superposition of Type 1 and Type 2 characteristics, but reflect an intrinsically distinct variability behavior. For $\sigma_{\mathrm{var}}$, the variability in the off-state sample is smaller than that in the on-state sample, but still larger than that of Type 2 AGNs. We also list the corresponding variability metric result of each sample in Table~\ref{tab:color_variability}, in which Type 2 AGNs exhibit significantly lower values, while the differences among the other samples remain relatively small.

We also performed KS tests on the distributions of variability metrics, and the results are shown in Table~\ref{tab:ks_test}. The KS statistics indicate that both the on-state and off-state CLAGNs differ substantially from the Type 1 and Type 2 AGN control samples, respectively. In contrast, the KS statistics are much smaller when comparing the on-/off-state CLAGNs with the overall CLAGNs, consistent with the similarities described above. These results quantitatively support our conclusions and are in agreement with the trends indicated by the higher fraction of sources with $\sigma_{\mathrm{QSO}} > 3$.

Physically, this behavior is consistent with the scenario in which CLAGNs experience significant changes in their accretion states \citep{Ricci2023}. The increase in variability compared to Type 2 AGNs suggests that at least part of the accretion disk becomes directly observable during the state transitions. Meanwhile, the similarity between CLAGNs and EVQs in their variability distributions implies that the same underlying mechanisms.

\section{Discussion} \label{sec:discussion}
\subsection{Variability - Type Transition?}\label{subsec:type transition}
Although the physical mechanisms driving CLAGNs remain unclear, variability studies play a crucial role in probing the physics of the central regions of AGNs. In particular, variability provides key constraints on accretion disk instabilities, the structure and reprocessing effects of the broad-line region (BLR), and the role of dust obscuration. Unlike the static spectral classification, variability-based studies offer a dynamic perspective on AGN evolution, making them essential for understanding the nature of CLAGNs. 

As shown in Table \ref{Table:photometry}, many previous searches for CLAGNs have focused on cases exhibiting extreme photometric variability. For these studies, the success rate of follow-up spectroscopic confirmation is highly dependent on the arbitrarily chosen variability threshold and the signal-to-noise ratio (SNR) of the observations, which remains moderate, typically around 15–50\% \citep{Yang2018, MacLeod2019}. While these selection methods have identified a number of CLAGNs, they may also introduce biases, potentially missing sources that undergo transitions in short timescales due to the time gaps between different surveys. The cadence and depth of photometric surveys also play a crucial role in detection efficiency, as shorter-duration surveys may fail to capture the full variability cycle of some CLAGNs.

In addition to photometric selection, some studies have employed variability-based methods to search for CLAGNs. \citet{Zhu2025} utilized a color-magnitude pattern as a selection criterion, achieving a success rate of approximately 45\%, comparable to the traditional photometric variability methods. \citet{WangShu2024} adopted a variability metric approach, yielding a higher success rate of around 80\%, demonstrating greater efficiency than the photometric methods. These studies generally assume that Type 1 and Type 2 AGNs exhibit distinct variability patterns, and classify CLAGNs by comparing their current variability properties to former spectral types. Implicitly, these methods rely on the assumption that the variability characteristics of CLAGNs are tied to their spectral classifications.

Our study specifically investigates the variability properties of CLAGNs and challenges the assumption that their variability mode is inherently tied to spectral classification. We find that variability-based transitions do not always correspond to spectral state transitions, suggesting that these two processes may be independent rather than intrinsically linked.

Traditionally, CLAGNs are defined by the appearance or disappearance of broad emission lines—a classification based on spectral features rather than a quantitative measure of variability. This definition implies a clear transition between a bright state and a dim state, each associated with a distinct AGN type. However, our analysis reveals that the variability properties of CLAGNs remain largely consistent across these spectral states. Additionally, CLAGNs exhibit variability characteristics that systematically differ from both Type 1 and Type 2 AGNs, challenging the notion that they undergo a complete and discrete type transition.

In our study, we found that CLAGNs exhibit an intermediate variability mode compared to Type 1 and Type 2 AGNs. A systematic comparison between on-state and off-state sources did not reveal significant differences, suggesting that the variability mode remains unchanged despite the observed spectral type transition. This result is consistent with \citet{Jana2025}, who compared the variability properties of CLAGNs and normal AGNs of different types (see their Figure 4). The existence of this intermediate variability mode raises important questions about the physical nature of CLAGNs—whether they undergo genuine transitions between Type 1 and Type 2 states or instead continuously occupy a critical regime between the two, with spectral transitions occurring as a result of moderate fluctuations in accretion rate. In this scenario, CLAGNs do not fully switch between distinct AGN types but rather oscillate within a transitional state, where changes in accretion conditions temporarily enhance or suppress broad-line emission. Further investigation into the variability timescales and Eddington ratio evolution of CLAGNs will be essential to distinguish between these possibilities and to establish whether accretion rate fluctuations alone can account for their observed spectral changes.

\subsection{Naked Type 2 AGN?}
The term naked Type 2 AGNs was first introduced to describe AGNs that lack both a broad-line region (BLR) and significant obscuration in the X-ray or optical bands \citep{Hawkins2004,Panessa2009}. Unlike traditional Type 2 AGNs, where the absence of broad emission lines is attributed to dust obscuration, naked Type 2 AGNs are thought to be intrinsically lacking a BLR, possibly due to a low accretion rate or a different accretion disk structure \citep{LaMassa2014}. Theoretical models suggest that at sufficiently low Eddington ratios ($R_{\mathrm{Edd}}\lesssim 0.01$), the accretion flow may become radiatively inefficient, suppressing the formation of a standard BLR\citep{Noda2018, Ricci2023}.

The nature of CLAGNs suggests a possible link to naked Type 2 AGNs. Since CLAGNs are once believed to exhibit transitions between Type 1 and Type 2 states, one scenario is that they represent AGNs fluctuating near the critical threshold where the BLR becomes unstable or disappears altogether. If CLAGNs in their Type 2 phase are intrinsically lacking a photoionized BLR rather than being obscured, they may effectively be transient naked Type 2 AGNs. This interpretation is supported by the fact that some CLAGNs remain in a Type 2 state without showing signs of heavy obscuration \citep{Yang2023, Jana2025}. In this case, CLAGNs could provide a valuable population for studying the physics of the BLR and its dependence on accretion rate.

Furthermore, if CLAGNs contain a significant fraction of naked Type 2 AGNs, they serve as ideal candidates for investigating variability in low-Eddington ratio AGNs, particularly in the X-ray regime. X-ray variability is a key diagnostic of accretion processes in AGNs, and in low-$\lambda_{\mathrm{Edd}}$ systems, changes in the spectral index ($\alpha_{\mathrm{OX}}$) have been shown to correlate with the Eddington ratio \citep{Yang2023,Hagen2024}. Systematically, when $R_{\mathrm{Edd}}\gtrsim0.01$, AGNs tend to exhibit harder X-ray spectra (lower $\alpha_{\mathrm{OX}}$) when $R_{\mathrm{Edd}}$ decrease. By contrast, when $R_{\mathrm{Edd}}\lesssim0.01$, $\alpha_{\mathrm{OX}}$ soften when $R_{\mathrm{Edd}}$ getting smaller, consistent with a transition to a radiatively inefficient accretion flow \citep{Yang2023}.

Tracking these trends across individual AGNs at different accretion rates, especially at the low-$R_{\mathrm{Edd}}$ end, is crucial for understanding accretion physics. However, variability studies of normal low-$R_{\mathrm{Edd}}$ AGNs are challenging, as they are often heavily obscured Type 2 AGNs, where intrinsic variability is significantly suppressed or entirely hidden. In contrast, CLAGNs, particularly those in their off-state, may provide a unique opportunity to probe accretion physics at low-$R_{\mathrm{Edd}}$ with less obscuration.

If CLAGNs in their off-state resemble low-$R_{\mathrm{Edd}}$ AGNs, their X-ray spectral variability can provide insight into the accretion physics at play \citep{Noda2018,Lyu2025_nustar}. Future studies combining X-ray data (e.g., from Chandra and XMM-Newton) and optical spectroscopic monitoring could test whether CLAGNs follow the same $\alpha_{\mathrm{OX}}$–$R_{\mathrm{Edd}}$ relation as other low-luminosity AGNs, further implying their nature.

\section{Conclusion} \label{sec:conclusion}
We constructed a sample of 152 CLAGNs from the literature, all exhibiting clear spectral transitions characterized by the appearance or disappearance of broad emission lines. To enable a meaningful comparison, we established a control sample consisting of EVQs, Type 1 AGNs, and Type 2 AGNs, matched in redshift and SMBH mass. This control sample allows us to investigate potential differences in variability behavior across various AGN populations.

We further divided the CLAGNs into turn-on and turn-off subsamples based on their most recent spectral classifications. This categorization enables a direct comparison between spectral transitions and corresponding photometric variability.

A bluer-when-brighter trend is confirmed for CLAGNs in the optical bands, as shown in their color–magnitude relation. CLAGNs occupy an intermediate region between Type 1 and Type 2 AGNs in this color–magnitude space. Despite their evident spectral transitions between on-state and off-state phases, we do not observe significant differences in their color–magnitude trends across these states.

In terms of structure function analysis, both CLAGNs and EVQs show systematically larger $A$ values than Type 1 and Type 2 AGNs, indicating stronger variability amplitudes. CLAGNs exhibit lower $\gamma$ values compared to EVQs and Type 1 AGNs, while Type 2 AGNs cluster at even lower $\gamma$.

We also investigated the variability metrics $\sigma_{\mathrm{QSO}}$ and $\sigma_{\mathrm{var}}$ across all samples. While most Type 1 AGNs show high values in both metrics and Type 2 AGNs cluster at low values, CLAGNs and EVQs display a continuous distribution centered in a critical regime between the two types. Notably, on-state CLAGNs do not strongly resemble typical Type 1 AGNs, nor do off-state CLAGNs align with Type 2 AGNs. Instead, both phases of CLAGNs share similar variability properties with each other, further supporting the idea that CLAGNs occupy a distinct and transitional variability regime.

Taken together, these results suggest that the variability pattern of CLAGNs does not fundamentally change with their spectral state. Rather than representing simple switches between classical Type 1 and Type 2 AGN modes, CLAGNs are more likely to represent a special, critical state of AGN activity.

\begin{acknowledgments}
We thank the anonymous referee for a very constructive report which significantly helps us to revise the paper. We thank the support of the National Key R\&D Program of China ( grant No. 2022YFF0503401) and the National Science Foundation of China ( grant No. 12133001). 
This work is based on observations obtained with the Samuel Oschin Telescope 48 inch and the 60 inch Telescope at the Palomar Observatory as part of the Zwicky Transient Facility project. ZTF is supported by the National Science Foundation under grant No. AST-2034437 and a
collaboration including Caltech, IPAC, the Weizmann Institute for Science, the Oskar Klein Center at Stockholm University, the University of Maryland, Deutsches Elektronen-Synchrotron and Humboldt University, the TANGO Consortium of Taiwan, the University of Wisconsin at Milwaukee, Trinity College Dublin, Lawrence Livermore National Laboratories, and IN2P3, France.
Operations are conducted by COO, IPAC, and UW.

Funding for the Sloan Digital Sky Survey IV has been provided by the Alfred P. Sloan Foundation, the U.S. Department of Energy Office of Science, and the Participating Institutions. 

SDSS-IV acknowledges support and resources from the Center for High Performance Computing  at the University of Utah. The SDSS website is www.sdss4.org.

SDSS-IV is managed by the 
Astrophysical Research Consortium 
for the Participating Institutions 
of the SDSS Collaboration including 
the Brazilian Participation Group, 
the Carnegie Institution for Science, 
Carnegie Mellon University, Center for 
Astrophysics | Harvard \& 
Smithsonian, the Chilean Participation 
Group, the French Participation Group, 
Instituto de Astrof\'isica de 
Canarias, The Johns Hopkins 
University, Kavli Institute for the 
Physics and Mathematics of the 
Universe (IPMU) / University of 
Tokyo, the Korean Participation Group, 
Lawrence Berkeley National Laboratory, 
Leibniz Institut f\"ur Astrophysik 
Potsdam (AIP),  Max-Planck-Institut 
f\"ur Astronomie (MPIA Heidelberg), 
Max-Planck-Institut f\"ur 
Astrophysik (MPA Garching), 
Max-Planck-Institut f\"ur 
Extraterrestrische Physik (MPE), 
National Astronomical Observatories of 
China, New Mexico State University, 
New York University, University of 
Notre Dame, Observat\'ario 
Nacional / MCTI, The Ohio State 
University, Pennsylvania State 
University, Shanghai 
Astronomical Observatory, United 
Kingdom Participation Group, 
Universidad Nacional Aut\'onoma 
de M\'exico, University of Arizona, 
University of Colorado Boulder, 
University of Oxford, University of 
Portsmouth, University of Utah, 
University of Virginia, University 
of Washington, University of 
Wisconsin, Vanderbilt University, 
and Yale University.

\end{acknowledgments}

\appendix
\section{Definition of variability metric}\label{Sec: Appendix}

We adopt two significance measures to assess the quasar-like variability in light curves, following the damped random walk (DRW) formalism described by \citet{Butler2011}.

For objects with $n$ light curve data points, the number of degrees of freedom $\nu=n-1$. Let the measured magnitudes at two epochs $i$ and $j$ are $m_i$ and $m_j$, their respective photometric uncertainties are $\sigma_i$ and $\sigma_j$, and the error-weights are, $w_i=1/\sigma_i^2$. From \citet{Butler2011}, the exponential damping timescale have, $\mathrm{log}(\tau_\circ) = 2.92-0.07\times(m-19)$, and the intrinsic variance between observations on short timescales $\tau_{ij} \sim 1\ \mathrm{days}$, $\mathrm{log}(\hat \sigma^2)=-4.1+0.14\times(m-19)$.

Supposed that the quasar variability as a function of time difference is modeled using a covariance matrix of the form:

\begin{equation}
C_{i j}=\sigma_i^2 \delta_{i j}+\frac{1}{2} \hat{\sigma}^2 \tau_{\circ} \exp \left(-\tau_{i j} / \tau_{\circ}\right)
\end{equation}
where $\delta_{ij}$ is the Kronecker delta function (1 for $i = j$, 0 otherwise)

Then we have,
\begin{equation}
\chi_{\mathrm{QSO}}^2=\left(x-x_{\circ, \mathrm{best}}\right)^T C^{-1}\left(x-x_{\circ, \mathrm{best}}\right),
\end{equation}
in which, $x_{\circ, \text { best }}=\sum_{i, j} C_{i j}^{-1} x_j / \sum_{i, j} C_{i j}^{-1}$.

The first metric, $\sigma_{\mathrm{QSO}}$, quantifies how significantly the observed variability resembles a quasar-like DRW model compared to a generic variable source. It is derived from the probability \( p_{\mathrm{QSO}} \) that the reduced chi-squared value of the DRW fit, \( \chi^2_{\mathrm{QSO}}/\nu \), is lower than that expected for a non-quasar variable. 

This probability is computed using the regularized incomplete beta function. Firstly, the ratio $x$ can be written as:
\begin{equation}
  x = \frac{\chi^2_{\text{QSO}}}{\chi^2_{\text{QSO}} + \mathrm{Tr}(C^{-1})\cdot v_m}, \mathrm{in\ which\ }v_m=\left\langle m^2\right\rangle-\langle m\rangle^2,
\end{equation}
then the probability could be calculated by the regularized incomplete beta function:
\begin{equation}
  p_{\mathrm{QSO}} = I_x(\frac{\nu}{2}, \frac{\nu}{2}),
\end{equation}
and the final significance is converted into Gaussian sigma units:
\begin{equation}
\sigma_{\mathrm{QSO}} = \Phi^{-1}\left(1 - \frac{1}{2} p_{\mathrm{QSO}}\right),
\end{equation}
where \( \Phi^{-1} \) is the inverse cumulative distribution function of the standard normal distribution.

The second metric, $\sigma_{\mathrm{var}}$, evaluates how strongly the light curve deviates from a constant-flux (non-variable) model. It is based on the classical reduced chi-squared value \( \chi_{\mathrm{var}}^2/\nu \), and the corresponding upper-tail probability \(p_{\mathrm{vary}} \), 

From the total chi-squared:

\begin{equation}
  x = \chi_{\mathrm{var}}^2= \sum_{i=1}^n \frac{\left(m_i-\bar{m}_w\right)^2}{\sigma_i^2},\mathrm{in\ which\ }\bar{m}_w=\frac{\sum_{i=1}^n w_i m_i}{\sum_{i=1}^n w_i}, 
\end{equation}
and the probability is described by the regularized upper incomplete gamma function:
\begin{equation}
  p_{\mathrm{vary}} = Q\left(\frac{\nu}{2}, \frac{x}{2}\right),
\end{equation}
and the final metric $\sigma_{\mathrm{var}}$ is computed from the regularized incomplete gamma function:
\begin{equation}
    \sigma_{\mathrm{var}} = \Phi^{-1}\left(1 - \frac{1}{2} p_{\mathrm{vary}}\right).
\end{equation}

Both metrics are expressed in units of Gaussian sigma, where higher values indicate stronger statistical significance. 

\begin{table}[htbp]
\centering
\caption{Summary of CLAGN Sources}\label{table: clagn}
\begin{tabular}{lccccc}
\hline
Reference & Name & Coordinate & Redshift & Phase  \\
\hline

\citet{MacLeod2016} & J0159+0033 & 01:59:57.64+00:33:10.50 & 0.31204 & off \\ 

& J0023+0035 & 00:23:11.06+00:35:17.53 & 0.422105 & on \\ 

& J0225+0030 & 02:25:56.08+00:30:26.72 & 0.503908 & on-off \\ 

& J0226-0039 & 02:26:52.24-00:39:16.50 & 0.625329 & off \\ 

& J1002+4509 & 10:02:20.19+45:09:27.30 & 0.400894 & off \\ 

& J1021+4645 & 10:21:52.35+46:45:15.71 & 0.20413 & off \\ 

& J1324+4802 & 13:24:57.30+48:02:41.32 & 0.27156 & off \\ 

& J2146+0009 & 21:46:13.31+00:09:30.79 & 0.622 & on \\ 

& J2252+0109 & 22:52:40.37+01:09:58.71 & 0.533515 & on \\ 

& J2333-0023 & 23:33:17.39-00:23:03.53 & 0.51301 & on \\ 

\hline

\citet{Yang2018} & J0831+3646 & 08:31:32.25+36:46:17.27 & 0.19501 & on  \\ 

& J0849+2747 & 08:49:57.79+27:47:28.96 & 0.29854 & off  \\ 

& J0909+4747 & 09:09:32.02+47:47:30.69 & 0.11694 & on \\

& J0937+2602 & 09:37:30.33+26:02:32.14 & 0.16219 & on \\ 

& J1003+3525 & 10:03:23.46+35:25:03.84 & 0.11886 & on \\

& J1104+6343 & 11:04:23.21+63:43:05.36 & 0.16427 & off \\ 

& J1104+0118 & 11:04:55.17+01:18:56.64 & 0.57514 & off \\

& J1110-0003 & 11:10:25.45-00:03:34.14 & 0.21922 & on \\ 

& J1115+0544 & 11:15:36.57+05:44:49.73 & 0.08995 & on \\ 

& J1118+3203 & 11:18:29.64+32:04:00.00 & 0.3651 & off \\ 

& J1132+0357 & 11:32:29.14+03:57:29.08 & 0.09089 & on \\ 

& J1150+3632 & 11:50:39.32+36:32:58.43 & 0.34004 & off \\ 

& J1152+3209 & 11:52:27.49+32:09:59.45 & 0.37432 & off \\ 

& J1259+5515 & 12:59:16.74+55:15:07.15 & 0.19865 & on \\ 

& J1319+6753 & 13:19:30.74+67:53:55.37 & 0.16643 & on \\ 

& J1358+4934 & 13:58:55.83+49:34:14.11 & 0.11592 & on \\ 

& J1447+2833 & 14:47:54.24+28:33:24.10 & 0.16344 & on \\ 

& J1533+0110 & 15:33:56.00+01:10:29.79 & 0.14268 & on\\ 

& J1545+2511 & 15:45:29.64+25:11:27.87 & 0.11696 & on \\ 

& J1550+4139 & 15:50:17.23+41:39:02.29 & 0.22014 & on\\

& J1552+2737 & 15:52:58.30+27:37:28.46 & 0.08648 & on \\ 

& J0126-0839 & 01:26:48.09-08:39:48.05 & 0.19791 & off \\ 

& J0159+0033 & 01:59:57.64+00:33:10.50 & 0.31204 & off \\ 

& J1011+5442 & 10:11:52.99+54:42:06.43 & 0.24639 & off\\ 

& J1554+3629 & 15:54:40.26+36:29:52.01 & 0.23683 & on \\ 

& J2336+0017 & 23:36:02.98+00:17:28.78 & 0.24283 & off \\ 

\hline

\citet{MacLeod2019} & J0009-1034 & 00:09:04.54-10:34:28.70 & 0.241 & off \\ 

& J0043+1344 & 00:43:39.33+13:44:36.55 & 0.527 & off \\ 

& J0134-0914 & 01:34:58.36-09:14:35.42 & 0.443 & off \\ 

& J0745+3809 & 07:45:11.98+38:09:11.32 & 0.236 & off \\ 

& J0927+0433 & 09:27:02.30+04:33:08.18 & 0.322 & off \\ 

& J1113+5313 & 11:13:29.68+53:13:38.78 & 0.239 & off \\ 

& J1233+0842 & 12:33:59.13+08:42:11.64 & 0.25488 & off\\ 

& J1434+5723 & 14:34:55.31+57:23:45.10 & 0.174 & off \\ 

& J1536+0342 & 15:36:12.80+03:42:45.85 & 0.365 & off \\ 

& J1537+4613 & 15:37:34.07+46:13:58.95 & 0.378 & off \\ 

& J1601+4745 & 16:01:11.26+47:45:09.69 & 0.297 & off \\ 

& J1617+0638 & 16:17:11.42+06:38:33.50 & 0.229 & off \\ 

& J1617+0638 & 16:17:11.42+06:38:33.50 & 0.229 & off\\ 

& J2102+0005 & 21:02:00.43+00:05:01.92 & 0.329 & off \\ 

\hline
\end{tabular}
\end{table}

\begin{table}[htbp]
\centering
\caption{Summary of CLAGN Sources (Continued)}
\begin{tabular}{lccccc}
\hline
Reference & Name & Coordinate & Redshift & Phase \\
\hline

\citet{MacLeod2019} & J2205-0711 & 22:05:37.72-07:11:14.58 & 0.295 & off \\ 

& J2252+0109 & 22:52:40.37+01:09:58.71 & 0.533515 & on-off \\ 

& J2351-0913 & 23:51:07.44-09:13:17.90 & 0.354 & off \\

\hline

\citet{Frederick2019} & J0817+1012 & 08:17:26.42+10:12:10.14 & 0.045841 & LINER-quasar \\ 

& J0915+4814 & 09:15:31.53+48:14:12.26 & 0.100518 & LINER-quasar \\ 

& J1133+6701 & 11:33:55.95+67:01:07.09 & 0.039757 & LINER-Sy1 \\ 

& J1225+5108 & 12:25:50.31+51:08:46.38 & 0.168017 & LINER-quasar \\ 

& J1254+4914 & 12:54:03.78+49:14:52.84 & 0.066947 & LINER-quasar \\ 

& J1533+4432 & 15:33:08.01+44:32:08.40 & 0.036743 & LINER-NLS1 \\ 

\hline
\citet{Sheng2019} & J1252+5918 & 12:52:58.72+59:18:32.76 & 0.124 & off; also WISE1\&WISE2 off \\ 

& J1307+4506 & 13:07:16.99+45:06:45.28 & 0.084 &off; also WISE1\&WISE2 off \\ 

& J1317+1024 & 13:17:37.92+10:24:27.81 & 0.28 & off; also WISE1\&WISE2 off \\

& J1428+1723 & 14:28:46.71+17:23:53.11 & 0.104 & off; also WISE1\&WISE2 off \\ 

& J1627+5419 & 16:27:52.19+54:19:12.48 & 0.316 & off; also WISE1\&WISE2 off \\ 

& J1713+2736 & 17:13:53.85+27:36:27.03 & 0.298 & off; also WISE1\&WISE2 off \\ 
\hline

\citet{PottsVillforth2021} & J0823+4220 & 08:23:23.89+42:20:48.30 & 0.152 & off \\ 

& J1723+5504 & 17:23:22.31+55:04:13.80 & 0.295 & on \\ 

& J0829+4154 & 08:29:42.67+41:54:36.90 & 0.126 & on \\ 

& J0002-0027 & 00:02:36.25-00:27:24.80 & 0.291 & off-on \\ 

\hline

\citet{Lopez2022} & J0010+0008 & 00:10:14.89+00:08:20.83 & 0.1022 & on \\ 

& J0113+0135 & 01:13:11.82+01:35:42.58 & 0.2375 & on \\ 

& J0755+1923 & 07:55:44.37+19:23:36.31 & 0.1083 & on \\ 

& J0812+0715 & 08:12:40.75+07:15:28.48 & 0.0849 & on \\ 

\hline

\citet{Lopez2023} & J1341+3700 & 13:41:48.78+37:00:47.13 & 0.2 & on \\ 

& J1203+6053 & 12:03:49.21+60:53:17.45 & 0.07 & on \\ 

& J1120+3418 & 11:20:08.69+34:18:45.83 & 0.04 & on \\ 

& J0832+3551 & 08:32:24.29+35:51:35.92 & 0.14 & on \\ 

& J0819+3019 & 08:19:17.51+30:19:35.76 & 0.1 & on \\ 

& J1200+1458 & 12:00:45.49+14:58:03.67 & 0.11 & on \\ 

& J1329-0130 & 13:29:35.86-01:30:34.03 & 0.08 & on \\ 

& J1538+4607 & 15:38:32.66+46:07:35.01 & 0.2 & on \\ 

& J1552+2102 & 15:52:59.94+21:02:46.89 & 0.17 & on \\ 

\hline

\citet{Yang2023} & J0206-0609 & 02:06:21.67-06:09:52.56 & 0.413 & off \\ 

& J0224-0918 & 02:24:29.10-09:18:51.70 & 0.357 & off \\ 

& J0829+4202 & 08:29:05.98+42:02:04.20 & 0.638 & off \\ 

& J1226-0011 & 12:26:38.66-00:11:13.99 & 0.642 & off \\ 

\hline

\citet{Zeltyn2024} & J0007+0058 & 00:07:50.51+00:58:15.20 & 0.345113 & on \\ 

& J0011+0100 & 00:11:03.49+01:00:32.62 & 0.485892 & off \\ 

& J0038+0002 & 00:38:41.53+00:02:26.70 & 0.649649 & off \\ 

&J0044-0106 & 00:44:59.09-01:06:29.30 & 0.228215 & off \\ 

&J0046+0002 & 00:46:16.68+00:02:49.09 & 0.4531 & on \\ 

&J0115+0033 & 01:15:36.11+00:33:52.42 & 0.364029 & on \\ 

&J0122-0002 & 01:22:56.19-00:02:52.58 & 0.340502 & on \\ 

&J0125+0019 & 01:25:24.66+00:19:35.90 & 1.662356 & off \\ 

&J0129+1504 & 01:29:46.72+15:04:57.29 & 0.364851 & off \\ 

&J0139-0042 & 01:39:44.05-00:42:51.52 & 0.471326 & off \\ 

&J0141+0037 & 01:41:28.85+00:37:12.22 & 0.814749 & off \\ 

&J0142-0050 & 01:42:09.02-00:50:49.99 & 0.132549 & on \\ 

&J0151-0034 & 01:51:05.80-00:34:26.40 & 0.335108 & off \\
& J0159+0033 & 01:59:57.64+00:33:10.51 & 0.31205 & off \\
\hline
\end{tabular}
\end{table}

\begin{table}[htbp]
\centering
\caption{Summary of CLAGN Sources (Continued)}
\begin{tabular}{lccccc}
\hline
Reference & Name & Coordinate & Redshift & Phase \\
\hline
\citet{Zeltyn2024} &J0206-0414 & 02:06:49.48-04:14:52.69 & 0.138743 & off \\ 

&J0221-0441 & 02:21:05.64-04:41:01.50 & 0.198636 & on \\ 

&J0239-0044 & 02:39:04.10-00:44:10.79 & 0.567837 & off \\ 

&J0245+0037 & 02:45:08.68+00:37:10.70 & 0.29935 & off \\ 

&J0759+3221 & 07:59:34.95+32:21:43.31 & 0.26888 & on \\ 

&J0801+3417 & 08:01:33.29+34:17:37.61 & 0.337982 & off \\ 

&J0807+1944 & 08:07:11.72+19:44:39.70 & 0.347745 & off \\ 

&J0807+2517 & 08:07:17.58+25:17:58.70 & 0.383252 & off \\ 

&J0810+2116 & 08:10:09.62+21:16:25.79 & 0.769466 & off \\ 

&J0813+2153 & 08:13:38.47+21:53:49.88 & 0.548448 & off \\ 

&J0815+3843 & 08:15:37.56+38:43:03.22 & 0.506738 & on \\ 

&J0820+3441 & 08:20:18.50+34:41:40.99 & 0.497461 & off \\ 

&J0834+2701 & 08:34:14.18+27:01:01.60 & 1.071886 & off \\ 

&J0841+0358 & 08:41:58.34+03:58:51.71 & 0.696807 & on \\ 

&J0845-0027 & 08:45:28.72-00:27:23.29 & 0.154388 & off \\ 

&J0846+0000 & 08:46:37.65+00:00:24.52 & 0.257489 & off \\ 

&J0849+0020 & 08:49:24.70+00:20:49.70 & 0.508254 & on \\ 

&J0853+0143 & 08:53:37.26+01:43:03.61 & 0.058189 & off \\ 

&J0855+0329 & 08:55:12.13+03:29:50.60 & 0.124046 & off \\ 

&J0902+0154 & 09:02:54.12+01:54:29.09 & 0.166121 & on \\ 

&J0903+0106 & 09:03:15.11+01:06:23.11 & 0.121657 & off \\ 

&J0905-0040 & 09:05:35.77-00:40:39.90 & 0.793011 & off \\ 

&J0905+0039 & 09:05:50.30+00:39:48.10 & 0.354041 & off \\ 

&J0906+0144 & 09:06:12.02+01:44:12.01 & 0.588482 & on \\ 

&J0908+0021 & 09:08:26.74+00:21:37.40 & 0.162574 & on \\ 

&J0916+0000 & 09:16:49.41+00:00:31.61 & 0.222549 & off \\ 

&J0923+0434 & 09:23:13.53+04:34:45.01 & 0.65682 & on \\ 

&J0926-0006 & 09:26:29.23-00:06:05.69 & 0.201324 & on-off \\ 

&J0927+0503 & 09:27:39.77+05:03:12.60 & 0.126061 & off \\ 

&J0929+0015 & 09:29:57.88+00:15:08.71 & 0.800271 & off \\ 

&J0930+0118 & 09:30:41.65+01:18:42.08 & 2.396249 & off \\ 

&J0932+0403 & 09:32:09.96+04:03:19.91 & 0.275805 & off \\ 

&J0933+0101 & 09:33:22.83+01:01:20.89 & 0.159933 & off \\ 

&J0936+0121 & 09:36:14.70+01:21:05.90 & 0.302738 & off \\ 

&J1037+3633 & 10:37:23.72+36:33:41.62 & 0.519399 & off \\ 

&J1125+3039 & 11:25:36.33+30:39:15.91 & 0.575978 & on \\ 

&J1128+5656 & 11:28:25.59+56:56:46.10 & 0.615026 & on \\ 

&J1130+3040 & 11:30:25.21+30:40:33.71 & 0.842421 & off \\ 

&J1131+5818 & 11:31:18.43+58:18:41.80 & 0.311977 & off \\ 

&J1133+5921 & 11:33:33.82+59:21:52.49 & 0.486347 & on \\ 

&J1216+5159 & 12:16:18.00+51:59:46.10 & 0.623324 & off \\ 

&J1324+4802 & 13:24:57.29+48:02:41.32 & 0.271847 & off \\ 

&J1354+5159 & 13:54:15.55+51:59:25.91 & 0.319749 & off \\ 

&J1357+5537 & 13:57:26.62+55:37:36.01 & 2.363621 & on \\ 

&J1357+6143 & 13:57:50.26+61:43:09.01 & 0.737404 & off \\ 

&J1405+5424 & 14:05:15.59+54:24:58.00 & 0.083262 & on \\ 

&J1411+5315 & 14:11:44.12+53:15:08.60 & 0.923187 & off \\ 

&J1426+3212 & 14:26:13.05+32:12:12.31 & 0.523328 & off \\ 

&J1428+0452 & 14:28:13.06+04:52:47.60 & 0.445953 & off \\
& J1440+3456 & 14:40:24.60+34:56:24.22 & 0.752449 & off \\
\hline
\end{tabular}
\end{table}

\begin{table}[htbp]
\centering
\caption{Summary of CLAGN Sources (Continued)}
\begin{tabular}{lccccc}
\hline
Reference & Name & Coordinate & Redshift & Phase \\
\hline
\citet{Zeltyn2024} &J1607+5603 & 16:07:30.23+56:03:05.62 & 0.717069 & on \\ 

&J1617+4902 & 16:17:34.42+49:02:11.90 & 0.880448 & off \\ 

&J1628+4329 & 16:28:29.18+43:29:48.59 & 0.260285 & off \\ 

&J1631+3844 & 16:31:10.02+38:44:57.41 & 0.583915 & off \\ 

&J1652+3937 & 16:52:34.96+39:37:33.38 & 0.641484 & on \\ 

&J2118+0053 & 21:18:34.39+00:53:22.20 & 0.451338 & on \\ 

&J2223+1944 & 22:23:06.32+19:44:31.60 & 0.259936 & off \\ 

&J2336+0040 & 23:36:50.69+00:40:42.49 & 0.161514 & off \\ 

\hline

\citet{Yang2024}&J0040+1609 & 00:40:38.40+16:09:49.93 & 0.2937904 & on \\ 

&J0047+1541 & 00:47:30.34+15:41:49.41 & 0.031420413 & on \\ 

&J0107+2428 & 01:07:47.92+24:28:48.70 & 0.1602249 & on \\ 

&J0110+0026 & 01:10:59.31+00:26:01.14 & 0.018818583 & on \\ 

&J0132+1501 & 01:32:42.04+15:01:14.76 & 0.17000946 & on \\ 

&J0141+0105 & 01:41:53.98+01:05:05.50 & 0.10127 & on \\ 

&J0146+1311 & 01:46:38.87+13:11:09.39 & 0.16013657 & on \\ 

&J0334+0051 & 03:34:31.32+00:51:21.37 & 0.4293435 & on \\ 

&J0803+2207 & 08:03:04.74+22:07:34.03 & 0.12452307 & on \\

&J0813+4608 & 08:13:19.32+46:08:49.69 & 0.053826876 & on \\ 

&J0823+4202 & 08:23:58.66+42:02:59.21 & 0.12564875 & on \\ 

&J0829+2319 & 08:29:23.00+23:19:32.70 & 0.1442577 & on \\ 

&J0837+0356 & 08:37:16.47+03:56:06.06 & 0.06348244 & on \\ 

&J0854+1113 & 08:54:34.65+11:13:34.78 & 0.16718863 & on \\ 

&J0901+2907 & 09:01:15.02+29:07:34.39 & 0.2176674 & on \\ 

&J0906+4046 & 09:06:12.28+40:46:38.51 & 0.1669009 & on \\ 

&J0908+0755 & 09:08:14.07+07:55:24.27 & 0.14367613 & on \\ 

&J0910+1907 & 09:10:13.70+19:07:01.37 & 0.18770769 & on \\ 

&J0914+0502 & 09:14:59.15+05:02:43.42 & 0.14251585 & on \\ 

&J0926-0006 & 09:26:29.23-00:06:05.72 & 0.20132396 & on \\ 

&J0936+2726 & 09:36:33.56+27:26:04.34 & 0.10286305 & on \\ 

&J0947+5449 & 09:47:46.56+54:49:55.82 & 0.6195132 & on \\ 

&J0952+2229 & 09:52:17.39+22:29:25.76 & 0.08075754 & on \\ 

&J1000+0354 & 10:00:01.81+03:54:02.04 & 0.21543361 & on \\ 

&J1002+0303 & 10:02:07.03+03:03:27.67 & 0.023351775 & on \\ 

&J1003+0202 & 10:03:50.98+02:02:27.65 & 0.12469647 & on \\ 

&J1003+0458 & 10:03:40.39+04:58:20.70 & 0.12374884 & on \\ 

&J1009+3539 & 10:09:06.07+35:39:32.65 & 0.110079706 & on \\ 

&J1012+5559 & 10:12:48.93+55:59:18.44 & 0.12547755 & on \\ 

&J1017+0658 & 10:17:37.94+06:58:16.18 & 0.045342874 & on \\ 

&J1058+6338 & 10:58:53.56+63:38:41.09 & 0.20183021 & on \\ 

&J1102+5722 & 11:02:00.11+57:22:50.50 & 0.14184508 & on \\ 

&J1105+0516 & 11:05:29.70+05:16:49.23 & 0.091155544 & on \\ 

&J1110+0804 & 11:10:00.32+08:04:15.71 & 0.116119295 & on \\ 

&J1111+1401 & 11:11:37.58+14:01:18.15 & 0.11065613 & on \\ 

&J1126+0510 & 11:26:21.87+05:10:02.97 & 0.17030554 & on \\ 

&J1149+3351 & 11:49:26.45+33:51:55.43 & 0.08091733 & on \\ 

&J1155+0048 & 11:55:47.64+00:48:52.00 & 0.20884128 & on \\ 

&J1158+1003 & 11:58:18.01+10:03:22.55 & 0.06992021 & on \\ 

&J1159+0513 & 11:59:53.39+05:13:29.98 & 0.059330106 & on \\ 

&J1159+6545 & 11:59:37.89+65:45:36.56 & 0.1218518 & on \\
& J1226+4559 & 12:26:35.48+45:59:32.38 & 0.10874906 & on \\ 
\hline
\end{tabular}
\end{table}

\begin{table}[htbp]
\centering
\caption{Summary of CLAGN Sources (Continued)}
\begin{tabular}{lccccc}
\hline
Reference & Name & Coordinate & Redshift & Phase  \\
\hline
\citet{Yang2024} &J1231+3232 & 12:31:55.14+32:32:40.26 & 0.06536951 & on \\ 

&J1239+0739 & 12:39:24.74+07:39:03.91 & 0.13254617 & on \\ 

&J1311+0705 & 13:11:03.04+07:05:09.73 & 0.30051038 & on \\ 

&J1311+4300 & 13:11:50.35+43:00:25.34 & 0.28453356 & on \\ 

&J1327+4025 & 13:27:23.74+40:25:04.48 & 0.0777004 & on \\ 

&J1328+6227 & 13:28:11.58+62:27:43.16 & 0.09058887 & on \\ 

&J1340+4006 & 13:40:54.65+40:06:37.41 & 0.17076361 & on \\ 

&J1341-0049 & 13:41:05.98-00:49:02.62 & 0.17537512 & on \\ 

&J1433+4943 & 14:33:18.21+49:43:01.29 & 0.12459213 & on \\ 

&J1442+5558 & 14:42:27.58+55:58:46.37 & 0.07689327 & on \\ 

&J1611+1642 & 16:11:13.84+16:42:03.29 & 0.15476471 & on \\ 

&J1612+3113 & 16:12:04.26+31:13:07.86 & 0.1773024 & on \\ 

&J1612+3816 & 16:12:46.01+38:16:43.85 & 0.19936875 & on \\ 

&J1636+3930 & 16:36:51.32+39:30:07.37 & 0.1818733 & on \\ 

&J1638+4712 & 16:38:06.33+47:12:12.67 & 0.114344485 & on \\ 

&J1640+2819 & 16:40:08.44+28:19:53.31 & 0.18136162 & on \\ 

&J1642+4423 & 16:42:38.27+44:23:31.62 & 0.22664222 & on \\ 

&J1643+5000 & 16:43:47.13+50:00:50.38 & 0.060020957 & on \\ 

&J1652+3416 & 16:52:21.38+34:16:41.70 & 0.26308012 & on \\ 

&J2120-0056 & 21:20:39.75-00:56:44.29 & 0.25076804 & on \\ 

&J2150-0106 & 21:50:55.73-01:06:54.19 & 0.08790665 & on \\ 

&J2156+2907 & 21:56:43.84+29:07:57.96 & 0.091868415 & on \\ 

&J2159+0556 & 21:59:05.79+05:56:39.46 & 0.1109222 & on \\ 

&J2201+0352 & 22:01:50.84+03:52:11.54 & 0.135897 & on \\ 

&J2203+1124 & 22:03:49.23+11:24:32.99 & 0.18626842 & on \\ 

&J2320+1448 & 23:20:10.99+14:48:15.15 & 0.08222562 & on \\ 

&J2336+1514 & 23:36:13.90+15:14:56.32 & 0.14734927 & on \\ 

&J2348+0741 & 23:48:54.90+07:41:36.68 & 0.46374503 & on \\
\hline
\end{tabular}
\end{table}

\vspace{5mm}
\facilities{Sloan, ZTF}

\software{
astropy \citep{2013A&A...558A..33A,2018AJ....156..123A}, astroquery \citep{Ginsburg2019},  pandas \citep{2022zndo...3509134T}, \texttt{qso\_fit} \citep{Butler2011}, TOPCAT \citep{Taylor2005}
}
\bibliography{sample631}{}

\begin{thebibliography}{}
\expandafter\ifx\csname natexlab\endcsname\relax\def\natexlab#1{#1}\fi
\providecommand{\url}[1]{\href{#1}{#1}}
\providecommand{\dodoi}[1]{doi:~\href{http://doi.org/#1}{\nolinkurl{#1}}}
\providecommand{\doeprint}[1]{\href{http://ascl.net/#1}{\nolinkurl{http://ascl.net/#1}}}
\providecommand{\doarXiv}[1]{\href{https://arxiv.org/abs/#1}{\nolinkurl{https://arxiv.org/abs/#1}}}

\bibitem[{{Astropy Collaboration} {et~al.}(2013){Astropy Collaboration}, {Robitaille}, {Tollerud}, {Greenfield}, {Droettboom}, {Bray}, {Aldcroft}, {Davis}, {Ginsburg}, {Price-Whelan}, {Kerzendorf}, {Conley}, {Crighton}, {Barbary}, {Muna}, {Ferguson}, {Grollier}, {Parikh}, {Nair}, {Unther}, {Deil}, {Woillez}, {Conseil}, {Kramer}, {Turner}, {Singer}, {Fox}, {Weaver}, {Zabalza}, {Edwards}, {Azalee Bostroem}, {Burke}, {Casey}, {Crawford}, {Dencheva}, {Ely}, {Jenness}, {Labrie}, {Lim}, {Pierfederici}, {Pontzen}, {Ptak}, {Refsdal}, {Servillat}, \& {Streicher}}]{2013A&A...558A..33A}
{Astropy Collaboration}, {Robitaille}, T.~P., {Tollerud}, E.~J., {et~al.} 2013, \aap, 558, A33, \dodoi{10.1051/0004-6361/201322068}

\bibitem[{{Astropy Collaboration} {et~al.}(2018){Astropy Collaboration}, {Price-Whelan}, {Sip{\H{o}}cz}, {G{\"u}nther}, {Lim}, {Crawford}, {Conseil}, {Shupe}, {Craig}, {Dencheva}, {Ginsburg}, {VanderPlas}, {Bradley}, {P{\'e}rez-Su{\'a}rez}, {de Val-Borro}, {Aldcroft}, {Cruz}, {Robitaille}, {Tollerud}, {Ardelean}, {Babej}, {Bach}, {Bachetti}, {Bakanov}, {Bamford}, {Barentsen}, {Barmby}, {Baumbach}, {Berry}, {Biscani}, {Boquien}, {Bostroem}, {Bouma}, {Brammer}, {Bray}, {Breytenbach}, {Buddelmeijer}, {Burke}, {Calderone}, {Cano Rodr{\'\i}guez}, {Cara}, {Cardoso}, {Cheedella}, {Copin}, {Corrales}, {Crichton}, {D'Avella}, {Deil}, {Depagne}, {Dietrich}, {Donath}, {Droettboom}, {Earl}, {Erben}, {Fabbro}, {Ferreira}, {Finethy}, {Fox}, {Garrison}, {Gibbons}, {Goldstein}, {Gommers}, {Greco}, {Greenfield}, {Groener}, {Grollier}, {Hagen}, {Hirst}, {Homeier}, {Horton}, {Hosseinzadeh}, {Hu}, {Hunkeler}, {Ivezi{\'c}}, {Jain}, {Jenness}, {Kanarek}, {Kendrew}, {Kern}, {Kerzendorf}, {Khvalko}, {King}, {Kirkby}, {Kulkarni},
  {Kumar}, {Lee}, {Lenz}, {Littlefair}, {Ma}, {Macleod}, {Mastropietro}, {McCully}, {Montagnac}, {Morris}, {Mueller}, {Mumford}, {Muna}, {Murphy}, {Nelson}, {Nguyen}, {Ninan}, {N{\"o}the}, {Ogaz}, {Oh}, {Parejko}, {Parley}, {Pascual}, {Patil}, {Patil}, {Plunkett}, {Prochaska}, {Rastogi}, {Reddy Janga}, {Sabater}, {Sakurikar}, {Seifert}, {Sherbert}, {Sherwood-Taylor}, {Shih}, {Sick}, {Silbiger}, {Singanamalla}, {Singer}, {Sladen}, {Sooley}, {Sornarajah}, {Streicher}, {Teuben}, {Thomas}, {Tremblay}, {Turner}, {Terr{\'o}n}, {van Kerkwijk}, {de la Vega}, {Watkins}, {Weaver}, {Whitmore}, {Woillez}, {Zabalza}, \& {Astropy Contributors}}]{2018AJ....156..123A}
{Astropy Collaboration}, {Price-Whelan}, A.~M., {Sip{\H{o}}cz}, B.~M., {et~al.} 2018, \aj, 156, 123, \dodoi{10.3847/1538-3881/aabc4f}

\bibitem[{{Becker} {et~al.}(1995){Becker}, {White}, \& {Helfand}}]{Becker1995}
{Becker}, R.~H., {White}, R.~L., \& {Helfand}, D.~J. 1995, \apj, 450, 559, \dodoi{10.1086/176166}

\bibitem[{{Butler} \& {Bloom}(2011)}]{Butler2011}
{Butler}, N.~R., \& {Bloom}, J.~S. 2011, \aj, 141, 93, \dodoi{10.1088/0004-6256/141/3/93}

\bibitem[{{Calzetti}(2001)}]{Calzetti2001}
{Calzetti}, D. 2001, \pasp, 113, 1449, \dodoi{10.1086/324269}

\bibitem[{{Cui} {et~al.}(2012){Cui}, {Zhao}, {Chu}, {Li}, {Li}, {Zhang}, {Su}, {Yao}, {Wang}, {Xing}, {Li}, {Zhu}, {Wang}, {Gu}, {Luo}, {Xu}, {Zhang}, {Liu}, {Zhang}, {Yang}, {Cao}, {Chen}, {Chen}, {Chen}, {Chen}, {Chu}, {Feng}, {Gong}, {Hou}, {Hu}, {Hu}, {Hu}, {Jia}, {Jiang}, {Jiang}, {Jiang}, {Jin}, {Li}, {Li}, {Li}, {Liu}, {Liu}, {Lu}, {Mao}, {Men}, {Qi}, {Qi}, {Shi}, {Tang}, {Tao}, {Wang}, {Wang}, {Wang}, {Wang}, {Wang}, {Wang}, {Wang}, {Wang}, {Wang}, {Wang}, {Wang}, {Wang}, {Xu}, {Xu}, {Yang}, {Yu}, {Yuan}, {Yuan}, {Zhai}, {Zhang}, {Zhang}, {Zhang}, {Zhao}, {Zhou}, {Zhou}, {Zhu}, \& {Zou}}]{Cui2012}
{Cui}, X.-Q., {Zhao}, Y.-H., {Chu}, Y.-Q., {et~al.} 2012, Research in Astronomy and Astrophysics, 12, 1197, \dodoi{10.1088/1674-4527/12/9/003}

\bibitem[{{DESI Collaboration} {et~al.}(2025){DESI Collaboration}, {Abdul-Karim}, {Adame}, {Aguado}, {Aguilar}, {Ahlen}, {Alam}, {Aldering}, {Alexander}, {Alfarsy}, {Allen}, {Allende Prieto}, {Alves}, {Anand}, {Andrade}, {Armengaud}, {Avila}, {Aviles}, {Awan}, {Bailey}, {Baleato Lizancos}, {Ballester}, {Bault}, {Bautista}, {BenZvi}, {Beraldo e Silva}, {Bermejo-Climent}, {Beutler}, {Bianchi}, {Blake}, {Blum}, {Bolton}, {Bonici}, {Brieden}, {Brodzeller}, {Brooks}, {Buckley-Geer}, {Burtin}, {Canning}, {Carnero Rosell}, {Carr}, {Carrilho}, {Casas}, {Castander}, {Cereskaite}, {Cervantes-Cota}, {Chaussidon}, {Chaves-Montero}, {Chen}, {Chen}, {Claybaugh}, {Cole}, {Cooper}, {Cousinou}, {Cuceu}, {Davis}, {Dawson}, {de Belsunce}, {de la Cruz}, {de la Macorra}, {de Mattia}, {Deiosso}, {Della Costa}, {Demina}, {Demirbozan}, {DeRose}, {Dey}, {Dey}, {Ding}, {Ding}, {Doel}, {Douglass}, {Dowicz}, {Ebina}, {Edelstein}, {Eisenstein}, {Elbers}, {Emas}, {Escoffier}, {Fagrelius}, {Fan}, {Fanning}, {Fawcett},
  {Fern\textbackslash'andez-Garc\textbackslash'ia}, {Ferraro}, {Findlay}, {Font-Ribera}, {Forero-Romero}, {Forero-S\textbackslash'anchez}, {Frenk}, {G\textbackslash''ansicke}, {Galbany}, {Garc\textbackslash'ia-Bellido}, {Garcia-Quintero}, {Garrison}, {Gazta\textbackslash\raisebox{-0.5ex}\textasciitilde naga}, {Gil-Mar\textbackslash'in}, {Gnedin}, {Gontcho}, {Gonzalez-Morales}, {Gonzalez-Perez}, {Gordon}, {Graur}, {Green}, {Gruen}, {Gsponer}, {Guandalin}, {Gutierrez}, {Guy}, {Hahn}, {Han}, {Han}, {He}, {Herrera-Alcantar}, {Honscheid}, {Hou}, {Howlett}, {Huterer}, {Ir\textbackslash v\{s\}i\textbackslash v\{c\}}, {Ishak}, {Jacques}, {Jimenez}, {Jing}, {Joachimi}, {Joudaki}, {Joyce}, {Jullo}, {Juneau}, {Kara\textbackslash c\{c\}ayl\{\textbackslash i\}}, {Karim}, {Kehoe}, {Kent}, {Khederlarian}, {Kirkby}, {Kisner}, {Kitaura}, {Kizhuprakkat}, {Kong}, {Koposov}, {Kremin}, {Krolewski}, {Lahav}, {Lai}, {Lamman}, {Lan}, {Landriau}, {Lang}, {Lange}, {Lasker}, {Le Goff}, {Le Guillou}, {Leauthaud}, {Levi}, {Li}, {Li},
  {Lodha}, {Lokken}, {Luo}, {Magneville}, {Manera}, {Manser}, {Margala}, {Martini}, {Maus}, {McCullough}, {McDonald}, {Medina}, {Medina-Varela}, {Meisner}, {Mena-Fern\textbackslash'andez}, {Menegas}, {Mezcua}, {Miquel}, {Montero-Camacho}, {Moon}, {Moustakas}, {Mu\textbackslash\raisebox{-0.5ex}\textasciitilde noz-Guti\textbackslash'errez}, {Mu\textbackslash\raisebox{-0.5ex}\textasciitilde noz-Santos}, {Myers}, {Myles}, {Nadathur}, {Najita}, {Napolitano}, {Newman}, {Nikakhtar}, {Nikutta}, {Niz}, {Noriega}, {Padmanabhan}, {Paillas}, {Palanque-Delabrouille}, {Palmese}, {Pan}, {Pan}, {Parkinson}, {Peacock}, {Percival}, {P\textbackslash'erez-Fern\textbackslash'andez}, {P\textbackslash'erez-R\textbackslash`afols}, \& {Peterson}}]{DESI2025}
{DESI Collaboration}, {Abdul-Karim}, M., {Adame}, A.~G., {et~al.} 2025, arXiv e-prints, arXiv:2503.14745, \dodoi{10.48550/arXiv.2503.14745}

\bibitem[{{Dong} {et~al.}(2025){Dong}, {Zhang}, {Gu}, {Sun}, \& {Zheng}}]{Dong2025}
{Dong}, Q., {Zhang}, Z.-X., {Gu}, W.-M., {Sun}, M., \& {Zheng}, Y.-G. 2025, \apj, 986, 160, \dodoi{10.3847/1538-4357/add331}

\bibitem[{{Ferrarese}(2002)}]{Ferrarese2002}
{Ferrarese}, L. 2002, \apj, 578, 90, \dodoi{10.1086/342308}

\bibitem[{Flesch(2021)}]{flesch2021millionquasarsmilliquasv72}
Flesch, E.~W. 2021, The Million Quasars (Milliquas) v7.2 Catalogue, now with VLASS associations. The inclusion of SDSS-DR16Q quasars is detailed.
\newblock \doarXiv{2105.12985}

\bibitem[{{Frederick} {et~al.}(2019){Frederick}, {Gezari}, {Graham}, {Cenko}, {van Velzen}, {Stern}, {Blagorodnova}, {Kulkarni}, {Yan}, {De}, {Fremling}, {Hung}, {Kara}, {Shupe}, {Ward}, {Bellm}, {Dekany}, {Duev}, {Feindt}, {Giomi}, {Kupfer}, {Laher}, {Masci}, {Miller}, {Neill}, {Ngeow}, {Patterson}, {Porter}, {Rusholme}, {Sollerman}, \& {Walters}}]{Frederick2019}
{Frederick}, S., {Gezari}, S., {Graham}, M.~J., {et~al.} 2019, \apj, 883, 31, \dodoi{10.3847/1538-4357/ab3a38}

\bibitem[{{Ginsburg} {et~al.}(2019){Ginsburg}, {Sip{\H{o}}cz}, {Brasseur}, {Cowperthwaite}, {Craig}, {Deil}, {Guillochon}, {Guzman}, {Liedtke}, {Lian Lim}, {Lockhart}, {Mommert}, {Morris}, {Norman}, {Parikh}, {Persson}, {Robitaille}, {Segovia}, {Singer}, {Tollerud}, {de Val-Borro}, {Valtchanov}, {Woillez}, {Astroquery Collaboration}, \& {a subset of astropy Collaboration}}]{Ginsburg2019}
{Ginsburg}, A., {Sip{\H{o}}cz}, B.~M., {Brasseur}, C.~E., {et~al.} 2019, \aj, 157, 98, \dodoi{10.3847/1538-3881/aafc33}

\bibitem[{{Graham} {et~al.}(2020){Graham}, {Ross}, {Stern}, {Drake}, {McKernan}, {Ford}, {Djorgovski}, {Mahabal}, {Glikman}, {Larson}, \& {Christensen}}]{Graham2020}
{Graham}, M.~J., {Ross}, N.~P., {Stern}, D., {et~al.} 2020, \mnras, 491, 4925, \dodoi{10.1093/mnras/stz3244}

\bibitem[{{Guo} {et~al.}(2024){Guo}, {Zou}, {Fawcett}, {Canning}, {Juneau}, {Davis}, {Alexander}, {Jiang}, {Aguilar}, {Ahlen}, {Brooks}, {Claybaugh}, {de la Macorra}, {Doel}, {Fanning}, {Forero-Romero}, {Gontcho A Gontcho}, {Honscheid}, {Kisner}, {Kremin}, {Landriau}, {Meisner}, {Miquel}, {Moustakas}, {Nie}, {Pan}, {Poppett}, {Prada}, {Rezaie}, {Rossi}, {Siudek}, {Sanchez}, {Schubnell}, {Seo}, {Sui}, {Tarl{\'e}}, \& {Zhou}}]{Guo2024}
{Guo}, W.-J., {Zou}, H., {Fawcett}, V.~A., {et~al.} 2024, \apjs, 270, 26, \dodoi{10.3847/1538-4365/ad118a}

\bibitem[{{Guo} {et~al.}(2025){Guo}, {Zou}, {Greenwell}, {Alexander}, {Fawcett}, {Pan}, {Siudek}, {Aguilar}, {Ahlen}, {Brooks}, {Claybaugh}, {Dawson}, {de la Macorra}, {Doel}, {Font-Ribera}, {Gazta{\~n}aga}, {Gontcho A Gontcho}, {Gutierrez}, {Kehoe}, {Kisner}, {Landriau}, {Le Guillou}, {Manera}, {Meisner}, {Miquel}, {Moustakas}, {Prada}, {Rossi}, {Sanchez}, {Schubnell}, {Sprayberry}, {Sui}, {Tarl{\'e}}, {Weaver}, {Xiao}, \& {Zou}}]{Guo2025}
{Guo}, W.-J., {Zou}, H., {Greenwell}, C.~L., {et~al.} 2025, \apjs, 278, 28, \dodoi{10.3847/1538-4365/adc124}

\bibitem[{{Hagen} {et~al.}(2024){Hagen}, {Done}, {Silverman}, {Li}, {Liu}, {Ren}, {Buchner}, {Merloni}, {Nagao}, \& {Salvato}}]{Hagen2024}
{Hagen}, S., {Done}, C., {Silverman}, J.~D., {et~al.} 2024, \mnras, 534, 2803, \dodoi{10.1093/mnras/stae2272}

\bibitem[{{Hawkins}(2004)}]{Hawkins2004}
{Hawkins}, M.~R.~S. 2004, \aap, 424, 519, \dodoi{10.1051/0004-6361:20041127}

\bibitem[{{Hopkins} {et~al.}(2007){Hopkins}, {Richards}, \& {Hernquist}}]{Hopkins2007}
{Hopkins}, P.~F., {Richards}, G.~T., \& {Hernquist}, L. 2007, \apj, 654, 731, \dodoi{10.1086/509629}

\bibitem[{{Jana} {et~al.}(2025){Jana}, {Ricci}, {Temple}, {Chang}, {Shablovinskaya}, {Trakhtenbrot}, {Diaz}, {Ilic}, {Nandi}, \& {Koss}}]{Jana2025}
{Jana}, A., {Ricci}, C., {Temple}, M.~J., {et~al.} 2025, \aap, 693, A35, \dodoi{10.1051/0004-6361/202451058}

\bibitem[{{Jin} {et~al.}(2022){Jin}, {Wu}, \& {Feng}}]{Jin2022}
{Jin}, J.-J., {Wu}, X.-B., \& {Feng}, X.-T. 2022, \apj, 926, 184, \dodoi{10.3847/1538-4357/ac410c}

\bibitem[{{Kelly} {et~al.}(2009){Kelly}, {Bechtold}, \& {Siemiginowska}}]{Kelly2009}
{Kelly}, B.~C., {Bechtold}, J., \& {Siemiginowska}, A. 2009, \apj, 698, 895, \dodoi{10.1088/0004-637X/698/1/895}

\bibitem[{{Kim} {et~al.}(2024){Kim}, {Son}, \& {Ho}}]{Kim2024}
{Kim}, M., {Son}, S., \& {Ho}, L.~C. 2024, \aap, 689, A27, \dodoi{10.1051/0004-6361/202450413}

\bibitem[{{Kovacevic} {et~al.}(2025){Kovacevic}, {Dai}, {Yuk}, {J{\"a}rvel{\"a}}, {Yi}, {Vallely}, {Shappee}, {Shankar}, \& {Stanek}}]{Kovacevic2025}
{Kovacevic}, N., {Dai}, X., {Yuk}, H., {et~al.} 2025, \apj, 985, 177, \dodoi{10.3847/1538-4357/adcb40}

\bibitem[{{Koz{\l}owski}(2016)}]{Kozlowski2016}
{Koz{\l}owski}, S. 2016, \apj, 826, 118, \dodoi{10.3847/0004-637X/826/2/118}

\bibitem[{{LaMassa} {et~al.}(2014){LaMassa}, {Yaqoob}, {Ptak}, {Jia}, {Heckman}, {Gandhi}, \& {Meg Urry}}]{LaMassa2014}
{LaMassa}, S.~M., {Yaqoob}, T., {Ptak}, A.~F., {et~al.} 2014, \apj, 787, 61, \dodoi{10.1088/0004-637X/787/1/61}

\bibitem[{{LaMassa} {et~al.}(2015){LaMassa}, {Cales}, {Moran}, {Myers}, {Richards}, {Eracleous}, {Heckman}, {Gallo}, \& {Urry}}]{LaMassa2015}
{LaMassa}, S.~M., {Cales}, S., {Moran}, E.~C., {et~al.} 2015, \apj, 800, 144, \dodoi{10.1088/0004-637X/800/2/144}

\bibitem[{{L{\'o}pez-Navas} {et~al.}(2022){L{\'o}pez-Navas}, {Mart{\'\i}nez-Aldama}, {Bernal}, {S{\'a}nchez-S{\'a}ez}, {Ar{\'e}valo}, {Graham}, {Hern{\'a}ndez-Garc{\'\i}a}, {Lira}, \& {Rojas Lobos}}]{Lopez2022}
{L{\'o}pez-Navas}, E., {Mart{\'\i}nez-Aldama}, M.~L., {Bernal}, S., {et~al.} 2022, \mnras, 513, L57, \dodoi{10.1093/mnrasl/slac033}

\bibitem[{{L{\'o}pez-Navas} {et~al.}(2023){L{\'o}pez-Navas}, {S{\'a}nchez-S{\'a}ez}, {Ar{\'e}valo}, {Bernal}, {Graham}, {Hern{\'a}ndez-Garc{\'\i}a}, {Homan}, {Krumpe}, {Lamer}, {Lira}, {Mart{\'\i}nez-Aldama}, {Merloni}, {R{\'\i}os}, {Salvato}, {Stern}, \& {Tub{\'\i}n-Arenas}}]{Lopez2023}
{L{\'o}pez-Navas}, E., {S{\'a}nchez-S{\'a}ez}, P., {Ar{\'e}valo}, P., {et~al.} 2023, \mnras, 524, 188, \dodoi{10.1093/mnras/stad1893}

\bibitem[{{Lyke} {et~al.}(2020){Lyke}, {Higley}, {McLane}, {Schurhammer}, {Myers}, {Ross}, {Dawson}, {Chabanier}, {Martini}, {Busca}, {Mas des Bourboux}, {Salvato}, {Streblyanska}, {Zarrouk}, {Burtin}, {Anderson}, {Bautista}, {Bizyaev}, {Brandt}, {Brinkmann}, {Brownstein}, {Comparat}, {Green}, {de la Macorra}, {Mu{\~n}oz Guti{\'e}rrez}, {Hou}, {Newman}, {Palanque-Delabrouille}, {P{\^a}ris}, {Percival}, {Petitjean}, {Rich}, {Rossi}, {Schneider}, {Smith}, {Vivek}, \& {Weaver}}]{Lyke2020}
{Lyke}, B.~W., {Higley}, A.~N., {McLane}, J.~N., {et~al.} 2020, \apjs, 250, 8, \dodoi{10.3847/1538-4365/aba623}

\bibitem[{{Lyu} {et~al.}(2025){Lyu}, {Yan}, {Wu}, {Wu}, {Yu}, \& {Liu}}]{Lyu2025_nustar}
{Lyu}, B., {Yan}, Z., {Wu}, X.-b., {et~al.} 2025, \mnras, 537, 1099, \dodoi{10.1093/mnras/staf109}

\bibitem[{{MacLeod} {et~al.}(2012){MacLeod}, {Ivezi{\'c}}, {Sesar}, {de Vries}, {Kochanek}, {Kelly}, {Becker}, {Lupton}, {Hall}, {Richards}, {Anderson}, \& {Schneider}}]{MacLeod2012}
{MacLeod}, C.~L., {Ivezi{\'c}}, {\v{Z}}., {Sesar}, B., {et~al.} 2012, \apj, 753, 106, \dodoi{10.1088/0004-637X/753/2/106}

\bibitem[{{MacLeod} {et~al.}(2016){MacLeod}, {Ross}, {Lawrence}, {Goad}, {Horne}, {Burgett}, {Chambers}, {Flewelling}, {Hodapp}, {Kaiser}, {Magnier}, {Wainscoat}, \& {Waters}}]{MacLeod2016}
{MacLeod}, C.~L., {Ross}, N.~P., {Lawrence}, A., {et~al.} 2016, \mnras, 457, 389, \dodoi{10.1093/mnras/stv2997}

\bibitem[{{MacLeod} {et~al.}(2019){MacLeod}, {Green}, {Anderson}, {Bruce}, {Eracleous}, {Graham}, {Homan}, {Lawrence}, {LeBleu}, {Ross}, {Ruan}, {Runnoe}, {Stern}, {Burgett}, {Chambers}, {Kaiser}, {Magnier}, \& {Metcalfe}}]{MacLeod2019}
{MacLeod}, C.~L., {Green}, P.~J., {Anderson}, S.~F., {et~al.} 2019, \apj, 874, 8, \dodoi{10.3847/1538-4357/ab05e2}

\bibitem[{{Masci} {et~al.}(2019){Masci}, {Laher}, {Rusholme}, {Shupe}, {Groom}, {Surace}, {Jackson}, {Monkewitz}, {Beck}, {Flynn}, {Terek}, {Landry}, {Hacopians}, {Desai}, {Howell}, {Brooke}, {Imel}, {Wachter}, {Ye}, {Lin}, {Cenko}, {Cunningham}, {Rebbapragada}, {Bue}, {Miller}, {Mahabal}, {Bellm}, {Patterson}, {Juri{\'c}}, {Golkhou}, {Ofek}, {Walters}, {Graham}, {Kasliwal}, {Dekany}, {Kupfer}, {Burdge}, {Cannella}, {Barlow}, {Van Sistine}, {Giomi}, {Fremling}, {Blagorodnova}, {Levitan}, {Riddle}, {Smith}, {Helou}, {Prince}, \& {Kulkarni}}]{Masci2019}
{Masci}, F.~J., {Laher}, R.~R., {Rusholme}, B., {et~al.} 2019, \pasp, 131, 018003, \dodoi{10.1088/1538-3873/aae8ac}

\bibitem[{{Mereghetti} {et~al.}(2021){Mereghetti}, {Balman}, {Caballero-Garcia}, {Del Santo}, {Doroshenko}, {Erkut}, {Hanlon}, {Hoeflich}, {Markowitz}, {Osborne}, {Pian}, {Rivera Sandoval}, {Webb}, {Amati}, {Ambrosi}, {Beardmore}, {Blain}, {Bozzo}, {Burderi}, {Campana}, {Casella}, {D'A{\'\i}}, {D'Ammando}, {De Colle}, {Della Valle}, {De Martino}, {Di Salvo}, {Doyle}, {Esposito}, {Frontera}, {Gandhi}, {Ghisellini}, {Gotz}, {Grinberg}, {Guidorzi}, {Hudec}, {Iaria}, {Izzo}, {Jaisawal}, {Jonker}, {Kong}, {Krumpe}, {Kumar}, {Manousakis}, {Marino}, {Martin-Carrillo}, {Mignani}, {Miniutti}, {Mundell}, {Mukai}, {Nucita}, {O'Brien}, {Orlandini}, {Orio}, {Palazzi}, {Papitto}, {Pintore}, {Piranomonte}, {Porquet}, {Ricci}, {Riggio}, {Rigoselli}, {Rodriguez}, {Saha}, {Sanna}, {Santangelo}, {Saxton}, {Sidoli}, {Stiele}, {Tagliaferri}, {Tavecchio}, {Tiengo}, {Tsygankov}, {Turriziani}, {Wijnands}, {Zane}, \& {Zhang}}]{Mereghetti2021}
{Mereghetti}, S., {Balman}, S., {Caballero-Garcia}, M., {et~al.} 2021, Experimental Astronomy, 52, 309, \dodoi{10.1007/s10686-021-09809-6}

\bibitem[{{Noda} \& {Done}(2018)}]{Noda2018}
{Noda}, H., \& {Done}, C. 2018, \mnras, 480, 3898, \dodoi{10.1093/mnras/sty2032}

\bibitem[{{Panessa} {et~al.}(2009){Panessa}, {Carrera}, {Bianchi}, {Corral}, {Gastaldello}, {Barcons}, {Bassani}, {Matt}, \& {Monaco}}]{Panessa2009}
{Panessa}, F., {Carrera}, F.~J., {Bianchi}, S., {et~al.} 2009, \mnras, 398, 1951, \dodoi{10.1111/j.1365-2966.2009.15225.x}

\bibitem[{{P{\^a}ris} {et~al.}(2018){P{\^a}ris}, {Petitjean}, {Aubourg}, {Myers}, {Streblyanska}, {Lyke}, {Anderson}, {Armengaud}, {Bautista}, {Blanton}, {Blomqvist}, {Brinkmann}, {Brownstein}, {Brandt}, {Burtin}, {Dawson}, {de la Torre}, {Georgakakis}, {Gil-Mar{\'\i}n}, {Green}, {Hall}, {Kneib}, {LaMassa}, {Le Goff}, {MacLeod}, {Mariappan}, {McGreer}, {Merloni}, {Noterdaeme}, {Palanque-Delabrouille}, {Percival}, {Ross}, {Rossi}, {Schneider}, {Seo}, {Tojeiro}, {Weaver}, {Weijmans}, {Y{\`e}che}, {Zarrouk}, \& {Zhao}}]{Paris2018}
{P{\^a}ris}, I., {Petitjean}, P., {Aubourg}, {\'E}., {et~al.} 2018, \aap, 613, A51, \dodoi{10.1051/0004-6361/201732445}

\bibitem[{{Pennell} {et~al.}(2017){Pennell}, {Runnoe}, \& {Brotherton}}]{Pennell2017}
{Pennell}, A., {Runnoe}, J.~C., \& {Brotherton}, M.~S. 2017, \mnras, 468, 1433, \dodoi{10.1093/mnras/stx556}

\bibitem[{{Potts} \& {Villforth}(2021)}]{PottsVillforth2021}
{Potts}, B., \& {Villforth}, C. 2021, \aap, 650, A33, \dodoi{10.1051/0004-6361/202140597}

\bibitem[{{Ren} {et~al.}(2022){Ren}, {Wang}, {Cai}, \& {Guo}}]{Ren2022}
{Ren}, W., {Wang}, J., {Cai}, Z., \& {Guo}, H. 2022, \apj, 925, 50, \dodoi{10.3847/1538-4357/ac3828}

\bibitem[{{Ricci} \& {Trakhtenbrot}(2023)}]{Ricci2023}
{Ricci}, C., \& {Trakhtenbrot}, B. 2023, Nature Astronomy, 7, 1282, \dodoi{10.1038/s41550-023-02108-4}

\bibitem[{{Schmidt} {et~al.}(2010){Schmidt}, {Marshall}, {Rix}, {Jester}, {Hennawi}, \& {Dobler}}]{Schmidt2010}
{Schmidt}, K.~B., {Marshall}, P.~J., {Rix}, H.-W., {et~al.} 2010, \apj, 714, 1194, \dodoi{10.1088/0004-637X/714/2/1194}

\bibitem[{{Sheng} {et~al.}(2020){Sheng}, {Wang}, {Jiang}, {Ding}, {Cai}, {Guo}, {Sun}, {Dou}, \& {Yang}}]{Sheng2019}
{Sheng}, Z., {Wang}, T., {Jiang}, N., {et~al.} 2020, \apj, 889, 46, \dodoi{10.3847/1538-4357/ab5af9}

\bibitem[{{Tang} {et~al.}(2023){Tang}, {Wolf}, \& {Tonry}}]{Tang2023}
{Tang}, J.-J., {Wolf}, C., \& {Tonry}, J. 2023, Nature Astronomy, 7, 473, \dodoi{10.1038/s41550-022-01885-8}

\bibitem[{{Taylor}(2005)}]{Taylor2005}
{Taylor}, M.~B. 2005, in Astronomical Society of the Pacific Conference Series, Vol. 347, Astronomical Data Analysis Software and Systems XIV, ed. P.~{Shopbell}, M.~{Britton}, \& R.~{Ebert}, 29

\bibitem[{{The pandas development Team}(2024)}]{2022zndo...3509134T}
{The pandas development Team}. 2024, {pandas-dev/pandas: Pandas}, v2.2.3,  Zenodo, \dodoi{10.5281/zenodo.3509134}

\bibitem[{{Thomas} {et~al.}(2013){Thomas}, {Steele}, {Maraston}, {Johansson}, {Beifiori}, {Pforr}, {Str{\"o}mb{\"a}ck}, {Tremonti}, {Wake}, {Bizyaev}, {Bolton}, {Brewington}, {Brownstein}, {Comparat}, {Kneib}, {Malanushenko}, {Malanushenko}, {Oravetz}, {Pan}, {Parejko}, {Schneider}, {Shelden}, {Simmons}, {Snedden}, {Tanaka}, {Weaver}, \& {Yan}}]{Thomas2013}
{Thomas}, D., {Steele}, O., {Maraston}, C., {et~al.} 2013, \mnras, 431, 1383, \dodoi{10.1093/mnras/stt261}

\bibitem[{{Urry} \& {Padovani}(1995)}]{Urry1995}
{Urry}, C.~M., \& {Padovani}, P. 1995, \pasp, 107, 803, \dodoi{10.1086/133630}

\bibitem[{{van Velzen} {et~al.}(2020){van Velzen}, {Holoien}, {Onori}, {Hung}, \& {Arcavi}}]{vanVelzen2020}
{van Velzen}, S., {Holoien}, T. W.~S., {Onori}, F., {Hung}, T., \& {Arcavi}, I. 2020, \ssr, 216, 124, \dodoi{10.1007/s11214-020-00753-z}

\bibitem[{{Vanden Berk} {et~al.}(2004){Vanden Berk}, {Wilhite}, {Kron}, {Anderson}, {Brunner}, {Hall}, {Ivezi{\'c}}, {Richards}, {Schneider}, {York}, {Brinkmann}, {Lamb}, {Nichol}, \& {Schlegel}}]{VB2004}
{Vanden Berk}, D.~E., {Wilhite}, B.~C., {Kron}, R.~G., {et~al.} 2004, \apj, 601, 692, \dodoi{10.1086/380563}

\bibitem[{{Vestergaard} \& {Peterson}(2006)}]{Vestergaard2006}
{Vestergaard}, M., \& {Peterson}, B.~M. 2006, \apj, 641, 689, \dodoi{10.1086/500572}

\bibitem[{{Wang} {et~al.}(2024){Wang}, {Woo}, {Gallo}, {Guo}, {Son}, {Kong}, {Mandal}, {Cho}, {Kim}, \& {Shin}}]{WangShu2024}
{Wang}, S., {Woo}, J.-H., {Gallo}, E., {et~al.} 2024, \apj, 966, 128, \dodoi{10.3847/1538-4357/ad3049}

\bibitem[{{Wu} \& {Shen}(2022)}]{Shen2022}
{Wu}, Q., \& {Shen}, Y. 2022, \apjs, 263, 42, \dodoi{10.3847/1538-4365/ac9ead}

\bibitem[{{Yang} {et~al.}(2025){Yang}, {Green}, {Wu}, {Eracleous}, {Jiang}, \& {Fu}}]{Yang2024}
{Yang}, Q., {Green}, P.~J., {Wu}, X.-B., {et~al.} 2025, \apj, 980, 91, \dodoi{10.3847/1538-4357/ad94ed}

\bibitem[{{Yang} {et~al.}(2018){Yang}, {Wu}, {Fan}, {Jiang}, {McGreer}, {Shangguan}, {Yao}, {Wang}, {Joshi}, {Green}, {Wang}, {Feng}, {Fu}, {Yang}, \& {Liu}}]{Yang2018}
{Yang}, Q., {Wu}, X.-B., {Fan}, X., {et~al.} 2018, \apj, 862, 109, \dodoi{10.3847/1538-4357/aaca3a}

\bibitem[{{Yang} {et~al.}(2023){Yang}, {Green}, {MacLeod}, {Plotkin}, {Anderson}, {Bieryla}, {Civano}, {Eracleous}, {Graham}, {Ruan}, {Runnoe}, \& {Zhao}}]{Yang2023}
{Yang}, Q., {Green}, P.~J., {MacLeod}, C.~L., {et~al.} 2023, \apj, 953, 61, \dodoi{10.3847/1538-4357/acdedd}

\bibitem[{{Zeltyn} {et~al.}(2024){Zeltyn}, {Trakhtenbrot}, {Eracleous}, {Yang}, {Green}, {Anderson}, {LaMassa}, {Runnoe}, {Assef}, {Bauer}, {Brandt}, {Davis}, {Frederick}, {Fries}, {Graham}, {Grogin}, {Guolo}, {Hern{\'a}ndez-Garc{\'\i}a}, {Koekemoer}, {Krumpe}, {Liu}, {Mart{\'\i}nez-Aldama}, {Ricci}, {Schneider}, {Shen}, {{\'S}niegowska}, {Temple}, {Trump}, {Xue}, {Brownstein}, {Dwelly}, {Morrison}, {Bizyaev}, {Pan}, \& {Kollmeier}}]{Zeltyn2024}
{Zeltyn}, G., {Trakhtenbrot}, B., {Eracleous}, M., {et~al.} 2024, \apj, 966, 85, \dodoi{10.3847/1538-4357/ad2f30}

\bibitem[{{Zhu} {et~al.}(2025){Zhu}, {Wang}, {Devanand}, {Gupta}, {Dogra}, {Li}, {Zhang}, {Ji}, \& {Sun}}]{Zhu2025}
{Zhu}, L.-T., {Wang}, Z., {Devanand}, P.~U., {et~al.} 2025, \mnras, 536, 2715, \dodoi{10.1093/mnras/stae2774}

\end{thebibliography}
\bibliographystyle{aasjournal}

\end{document}